%% file: deltaOri_2_rev3.tex
\documentclass[11pt,preprint2]{aastex}  

\usepackage{xspace} \usepackage{amssymb} \usepackage{times}
\usepackage{verbatim} \usepackage{color} \usepackage{graphics,graphicx}

\newcommand{\mone}  {^{-1}}
 \newcommand{\mtwo}  {^{-2}}

\newcommand{ \ang}{{\AA}\xspace}
\newcommand{\cmmtwo}{\,\mathrm{cm\mtwo}}

\newcommand{\cts}{\,\mathrm{cts\,s\mone}}

\newcommand{\rsun}  {\,R_\odot}

 \newcommand{\obs}{{ObsID}\xspace}

\newcommand{\delOri} {{$\delta\,$Ori}\xspace}
\newcommand{\Msun}{M$_{\odot}$}
\newcommand{\Rsun}{R$_{\odot}$}
\newcommand{\chandra}{\textit{Chandra}}
\newcommand{\einstein}{\textit{Einstein}}

\newcommand{\nefir}{\ion{Ne}{9}}
\newcommand{\neH}{\ion{Ne}{10}}
\newcommand{\mgfir}{\ion{Mg}{11}}
\newcommand{\mgH}{\ion{Mg}{12}}
\newcommand{\sifir}{\ion{Si}{13}}
\newcommand{\siH}{\ion{Si}{14}}
\newcommand{\sfir}{\ion{S}{15}}
\newcommand{\sH}{\ion{S}{16}}

\newcommand{\ofir}{\ion{O}{7}}
\newcommand{\oH}{\ion{O}{8}}

\newcommand{\ironsevt}{\ion{Fe}{17}}

 \newcommand{\ironninet}{\ion{Fe}{19}}
\newcommand{\irontwenty}{\ion{Fe}{20}}
\newcommand{\ks}{km s$^{-1}$}

\newcommand{\del}{$\delta$}

\newcommand{\capion[2]}{\mbox{#1\hspace{0.2em}{\rmfamily\@Roman{#2}}}}

\shorttitle{$\delta\,$Ori Aa X-ray Variability} \shortauthors{Nichols et
al.}

\begin{document}

\title{A Coordinated X-ray and Optical Campaign of the Nearby Massive
Binary \del\ Orionis Aa: II.  X-ray Variability\footnote{Based on data
from the \chandra\ X-ray Observatory and the MOST satellite, a Canadian
Space Agency mission, jointly operated by Dynacon Inc., the University
of Toronto Institute of Aerospace Studies, and the University of British
Columbia, with the assistance of the University of Vienna.} }

\author{J. Nichols\altaffilmark{1},  D.~P.~Huenemoerder\altaffilmark{2},
 M.~ F.~ Corcoran\altaffilmark{3}, W.~Waldron\altaffilmark{4},
Y.~Naz\'e\altaffilmark{5}, A.~M.~T.~ Pollock\altaffilmark{6},   A.~ F.~
J.~ Moffat\altaffilmark{7},  J.~Lauer\altaffilmark{1},  T.~Shenar
\altaffilmark{8}, C.~M.~P.~ Russell\altaffilmark{15,16},
N.~D.~Richardson\altaffilmark{7}, H. ~Pablo\altaffilmark{7},
N.~R.~Evans\altaffilmark{1}, K.~ Hamaguchi\altaffilmark{3,9},
T.~Gull\altaffilmark{10}, W.-R.~Hamann \altaffilmark{8}, L.~ Oskinova
\altaffilmark{8},  R.~ Ignace \altaffilmark{11},
Jennifer~L.~Hoffman\altaffilmark{12},  K.~ T.~ Hole\altaffilmark{13},
and J.~R.~Lomax\altaffilmark{14} } \altaffiltext{1}{Harvard-Smithsonian
Center for Astrophysics, Cambridge, MA, USA}
\altaffiltext{2}{Kavli Institute for Astrophysics and Space Research,
MIT, Cambridge, MA, USA}
\altaffiltext{3}{ USRA CRESST, Universities Space Research Association,
GSFC}
\altaffiltext{4}{Eureka Scientific, Inc, 2452 Delmer St., Oakland, CA
94602}
\altaffiltext{5}{FNRS/Dept AGO, Univ. of Li\`ege, All\'ee du 6 Ao\^ut
19c B5C, 4000-Li\`ege, Belgium}
\altaffiltext{6}{European Space Agency, \textit{XMM-Newton} Science
Operations Centre, European Space Astronomy Centre, Apartado 78, 28691
Villanueva de la Ca\~{n}ada, Spain}
\altaffiltext{7}{D\'epartement de physique, Universit\'e de Montr\'eal,
C.P. 6128, Succ. C.-V., QC, H3C 3J7, Canada} \altaffiltext{8}{Institut
f\"ur Physik und Astronomie, Universit\"at Potsdam, Karl-Liebknecht-Str.
24/25, D-14476 Potsdam, Germany} \altaffiltext{9}{Department of Physics,
University of Maryland, Baltimore County, 1000 Hilltop Circle,
Baltimore, MD 21250, USA} \altaffiltext{10}{Code 667, NASA/GSFC,
Greenbelt, MD 20771 USA} \altaffiltext{11}{Physics \& Astronomy, East
Tennessee State University, Johnson City, TN 37614 USA}
\altaffiltext{12}{Department of Physics and Astronomy, University of
Denver, 2112 E. Wesley Ave., Denver, CO, 80208 USA}
\altaffiltext{13}{Department of Physics, Weber State University, 2508
University Circle, Ogden, UT 84408} \altaffiltext{14}{Homer L. Dodge
Department of Physics and Astronomy, University of Oklahoma, 440 W
Brooks St, Norman, OK, 73019 USA} \altaffiltext{15}{X-ray Astrophysics
Lab, Code 662, NASA Goddard Space Flight Center, Greenbelt, MD 20771
USA} \altaffiltext{16}{Oak Ridge Associated Universities (ORAU), Oak
Ridge, TN 37831 USA}

\begin{abstract} We present time-resolved and phase-resolved variability
studies  of an extensive X-ray high-resolution spectral dataset  of the
\delOri\ Aa binary system. The  four observations, obtained with
\chandra\ ACIS HETGS, have a total exposure time of $\approx$479 ks and
provide nearly complete binary phase coverage.    Variability of the
total X-ray flux in the range 5-25 \ang\ is confirmed, with maximum
amplitude of about $\pm$15\%  within a single $\approx$125 ks
observation.    Periods of 4.76d and 2.04d are found in the total X-ray
flux, as well as an apparent overall increase in flux level throughout
the 9-day observational campaign.  Using 40 ks contiguous spectra
derived from the original observations, we investigate variability of
emission line parameters and ratios.  Several emission lines are shown
to be variable, including \sfir, \sifir, and \nefir. For the first time,
variations of the X-ray emission line widths as a function of the binary
phase are found  in a binary system, with the smallest widths at
$\phi$=0.0 when the secondary \delOri\ Aa2 is at inferior conjunction.
Using 3D hydrodynamic modeling of the interacting winds, we relate the
emission line width variability to the presence of a wind cavity created
by a wind-wind collision, which is effectively void of embedded wind
shocks and is carved out of the X-ray-producing primary wind, thus
producing phase locked X-ray variability. \end{abstract}

\section{INTRODUCTION} \label{section:introduction} Stellar winds of hot
massive stars, primarily those with M $\ge$ 8 \Msun, have important
effects on stellar and galactic evolution.  These winds provide
enrichment to the local interstellar medium via stellar mass loss.  On a
larger scale, the cumulative enrichment and  energy from the collective
winds of massive stars in a galaxy are expected to play a pivotal role
in  driving galactic winds \citep{leitherer92,oppenheimer06,mckee07}.
The number of massive stars in any star-forming galaxy, as well as their
tendency to be found in clusters, are critical parameters for
determining a galaxy's energy budget and evolution. \citet{oskinova05}
and   \citet{agertz13} showed that the energy from the winds of massive
stars will dominate over the energy from supernovae in the early years
of massive star cluster evolution. While substantial progress has been
made over the last several decades in modeling massive star winds
\citep{puls96}, many questions remain, such as the degree of clumping of
the winds, the radial location of different ions and temperature regimes
in the wind with respect to the stellar surface, and the origin of
Corotating Interaction Regions (CIRs) representing large-scale wind
perturbations.  X-ray observations have provided powerful diagnostic
tools for testing models, but a fully consistent description of the
detailed structure of a stellar wind is still elusive.

Variability in the  winds  of massive stars can  be an important probe
of the structure of the stellar winds.  There can be multiple causes of
X-ray variability in massive stars. Large-scale structures in the winds,
as traced by Discrete Absorption Components \citep[DACs,][]{kaper99}  and
possibly linked to CIRs, may be associated with shocks in the wind and
thereby potentially affect the X-ray emission. X-ray variations of this
type have probably been detected for $\zeta$ Oph, $\zeta$ Pup, and $\xi$
Per  \citep{osk01,naze13,massa14}. Also, X-ray variability with the same
period as but larger amplitude than known pulsational activity in the
visible domain was recently detected in $\xi^1$ CMa \citep{osk14} and
possibly in the hard band of $\beta$ Cru \citep{cohen08}. The exact
mechanism giving rise to these changes remains unclear.  Notably,  other
pulsating massive stars do not show such X-ray ``pulsations," such as
$\beta$ Cen \citep{raassen05} and $\beta$ Cep \citep{favata09}.
Smaller-scale structures such as clumps can also produce X-rays, albeit
at lower energies than large-scale structures.  It is also possible that
some X-ray variations in massive stars are stochastic in nature and are
not correlated with any currently known timescale.

Another cause for X-ray variability is possible in magnetic stars.  When
a strong global magnetic field exists, the stellar wind is forced to
follow the field lines, and the wind flowing from the two stellar
hemispheres may then collide at the equator, generating X-rays
\citep{babel97}.  Such recurrent variations have been detected in
$\theta^1$ Ori C \citep{stelzer05,gagne05}, HD~191612 \citep{naze10},
and possibly Tr16-22 \citep{naze14}, although the absence of large
variations in the X-ray emission of the magnetic star $\tau$ Sco is
puzzling in this context \citep{ignace10}.  Even stronger but very
localized magnetic fields could also be present, e.g. associated with
bright spots on the stellar surface that are required to create CIRs
\citep{cranmer96,cantiello09}.

In multiple systems, the collision of the wind of one star with the wind
of another can produce X-ray variations \citep{stevens92}.  Wind-wind
collision emission may vary with binary phase, with  inverse distance in
eccentric systems, or due to changes in line of sight absorption, as
observed  in HD~93403 \citep{rauw02}, Cyg OB2 9  \citep{naze12}, V444
Cyg \citep{lomax15}, and possibly HD~93205 \citep{antokhin03}.

Delta Ori A (Mintaka, HD~36486, 34~Ori) is a nearby multiple system that
includes a close eclipsing binary, \del\ Ori~Aa1
\citep[O9.5~II:][]{walborn72} and \del\ Ori~Aa2 (\citep[B1~V:][]{shenar15}, herein Paper IV), with period $\approx$5.73d
\citep{harvin02,mayer10}.   This close binary is orbited by a more distant companion star,
\del\ Ori~Ab ($\approx$B0~IV: \citet{pablo15}, herein
Paper III; Paper IV),  having a period of $\approx$346 yr
\citep{heintz87,perryman97,tokovinin14}.  The  components Aa1 and Aa2
are separated by about 43 \Rsun (2.6 $R_{Aa1}$; Paper III), and the
inclination of  $\approx$76\degr$\pm$4\degr\  (Paper III) ensures
eclipses. We acquired $\approx$479 ks of high resolution X-ray grating
spectra with a \chandra\ Large Program to observe nearly a full period
of \delOri\ Aa (\citet{corcoran15}, herein Paper I).
Simultaneous with the acquisition of the \chandra\ data,
Microvariability and Oscillations of STars (MOST)  space-based
photometry and ground-based spectroscopy at numerous geographical
locations were obtained and are reported in Paper III.

\input{parameters.tex}

%
Previous X-ray observations of the \delOri\ A system  from \einstein\
showed no significant variability \citep{grady84}. ROSAT data for
\delOri\  A were studied by \citet{Haberl93}, who found modest
2-$\sigma$ variability but  no obvious phase dependence;  the
\citet{corcoran96} reanalysis of the ROSAT data showed similar results.
A single previous \chandra\ HETGS observation of \delOri\ Aa was
analyzed by \citet{miller02}. Fitting the emission lines using Gaussian
profiles,  they found the profiles to be symmetrical and of low FWHM,
considering the estimated wind velocity. The 60 ks exposure time  covers
about 12\% of the orbital period.  \citet{pollock13} analyzed a
\chandra\ LETGS observation with exposure time of 96 ks, finding some
variability in the zeroth order image; they were not able to detect any
variability in the emission lines between two time splits of the
observation.

This paper is part of a series of papers.  Other papers in this series
address the parameters of the composite \chandra\ $\approx$479 ks
spectrum (Paper I), the simultaneous MOST and spectroscopic observations
(Paper III), and UV-optical-X-ray wind modeling (Paper IV).  In this
paper (Paper II), we investigate variability in the X-ray flux in the
\chandra\ spectra.  Sect.\ \ref{section:observations} describes the
\chandra\ data and processing techniques.  Sect.\ \ref{section:lc}
discusses the overall X-ray flux variability of the observations and
period search.    Time-resolved and phase-resolved analyses of emission
lines are presented in Sect.\ \ref{section:data_analysis}. In Sect.\
\ref{section:discussion} we relate our results of phase-based variable
emission line widths to a colliding wind model developed for this binary
system in Paper I, and discuss possible additional sources of
variability in \delOri\ Aa.   Sect.\ \ref{section:conclusions} presents
our conclusions.

\section{OBSERVATIONS AND DATA REDUCTION} \label{section:observations}

Delta Ori Aa was observed with the \chandra\ Advanced CCD Imaging
Spectrometer (ACIS) instrument using the High Energy Transmission
Grating Spectrometer (HETGS) \citep{canizares05}  for a total exposure
time of $\approx$479 ks, covering parts of 3 binary periods (see Table
\ref{tab:obs} for a list of observations and binary phases).  Four
separate observations were obtained within a 9-day interval.  The HETGS
consists of 2 sets of gratings: the Medium Energy Grating (MEG) with a
range of 2.5-31  \ang\ (0.4-5.0 keV) and resolution of 0.023 \ang\ FWHM,
and the High Energy Grating (HEG) with a range of 1.2-15 \ang\ (0.8-10
keV) and resolution of 0.012 \ang\ FWHM. The resolution is approximately
independent of wavelength.  The \chandra\ ACIS detectors record both MEG
and HEG dispersed grating spectra as well as the zeroth order image.
Due to spacecraft power considerations,  it was necessary to use only 5
ACIS CCD chips instead of the requested 6 for these observations.  Chip
S5 was not used, meaning wavelengths longer than about 19 \ang\  in the
MEG dispersed spectra  and about 9.5 \ang\ in the HEG dispersed spectra
were only recorded for the ``plus'' side of the dispersed spectra,
reducing the number of counts and effective exposure in these wavelength
regions. The standard data products distributed by the \chandra\ X-ray
Observatory were further processed with TGCat
software\footnote{available for public download: http://tgcat.mit.edu}
\citep{Hu11}. Specifically, each level 1 event file was processed into a
new level 2 event file using a package of standard CIAO analysis tools
\citep{ciao06}.   Additionally, appropriate redistribution matrix (RMF)
and area auxiliary response (ARF) files were calculated for each order
of each spectrum. TGCat processing produced analysis products with
supplemental statistical information, such as broad- and narrow-band
count rates.

\input{obs_table.tex}

\begin{figure*}[!tb] \centering\leavevmode
\includegraphics[width=5.5in]{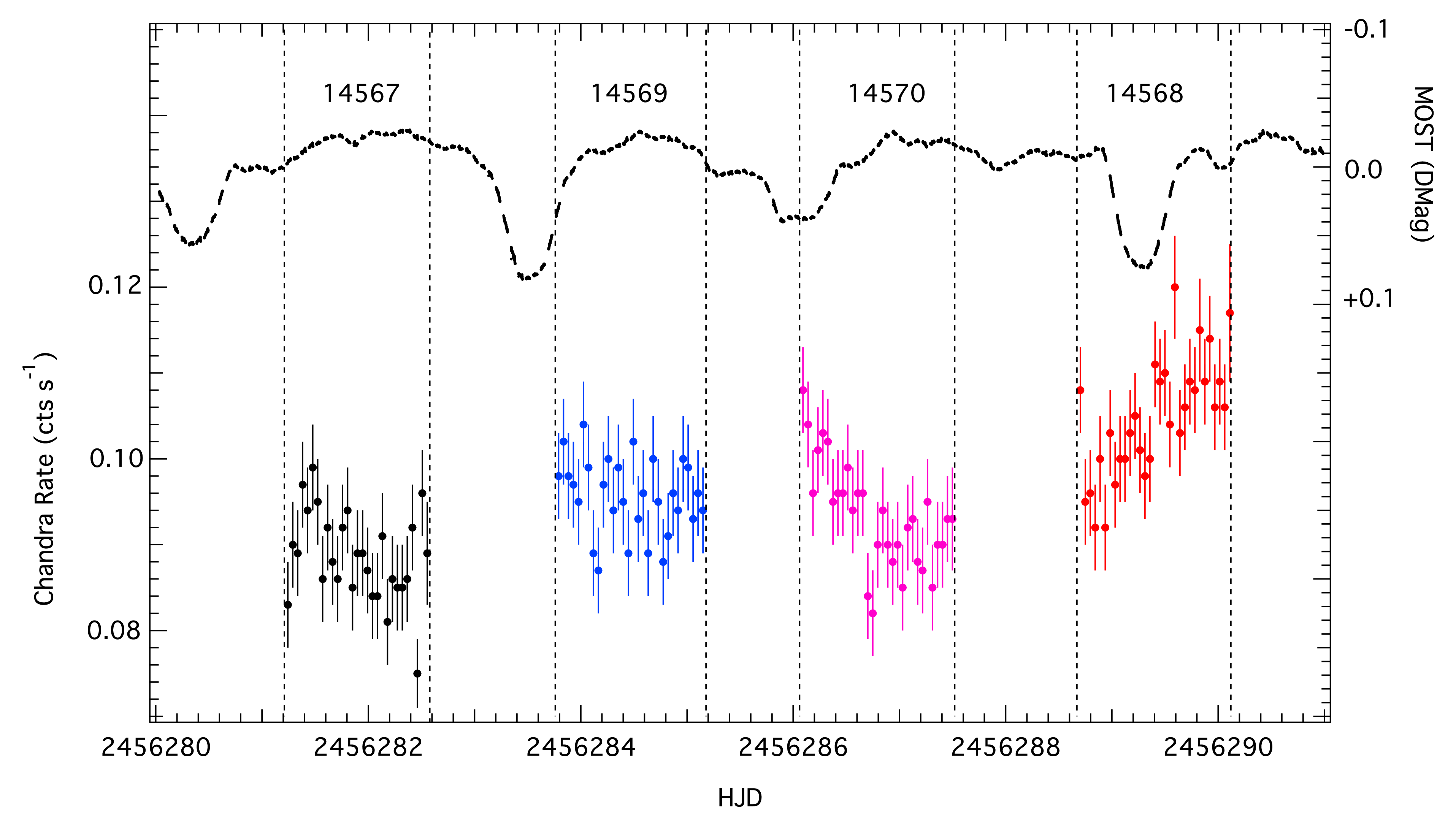} \caption{Chandra
X-ray lightcurve from the 2012 campaign with the simultaneous continuous
MOST optical lightcurve.  The time intervals for each of the Chandra observations are delineated with vertical lines with the Chandra observation ID (ObsID in Table \ref{tab:obs}) at the top of the figure.  The  Chandra lightcurves were
calculated from the dispersed spectra in each observation. The four
observations are separated by gaps due to the passage of Chandra through
the Earth$’$s radiation zone as well as necessary spacecraft thermal
control, during which time continued \delOri\ Aa observations were not
possible. Chandra counts per second are on the left y-axis. MOST
differential magnitudes are on the right y-axis.}\label{most_time}
\end{figure*}

\begin{figure*}[!tbh] \centering\leavevmode
\includegraphics[width=5.5in]{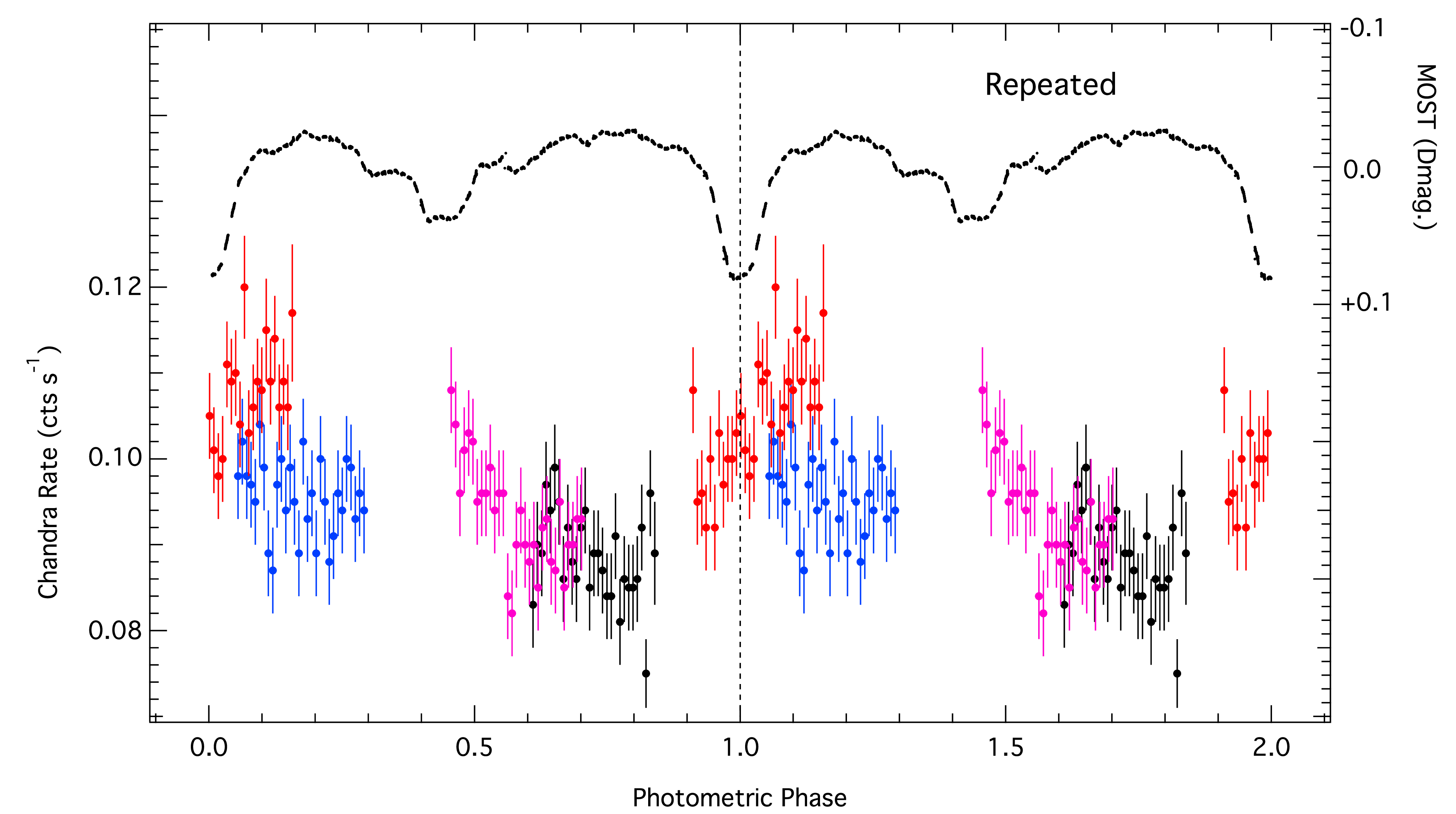} \caption{Phased
Chandra X-ray lightcurve from the 2012 campaign with the simultaneous
continuous MOST optical lightcurve. The mean \delOri\ Aa MOST optical
lightcurve is plotted above the Chandra X-ray lightcurve. The four
Chandra observations are shown in red, magenta, blue, and black. Binary
phase is on the x-axis, MOST differential magnitude on the right y-axis,
and Chandra count rate on the left y-axis. The MOST lightcurves have
been smoothed and both lightcurves have been repeated for one binary
orbit for clarity.} \label{most} \end{figure*}

MOST photometry observations were obtained of \delOri\ Aa for
approximately 3 weeks, including the 9 days of the \chandra\
observations.    Fig.\ \ref{most_time} shows the  simultaneous MOST
lightcurve aligned in time to the \chandra\ lightcurve.   The \chandra\
lightcurve in this figure is the $\pm$1 orders of the HEG and MEG
combined in the $1.7 \le \lambda \le 25.0$ \ang\ range, binned at 4 ks,
with Poisson errors.  Fig.\ \ref{most} shows the same data plotted with
binary phase rather than time.  The MOST lightcurves from several orbits
are averaged and overplotted in the figure to show the optical
variability.

Table \ref{table:params} lists the spectral types and radii of the \del\ Ori Aa1 and Aa2, as well as the orbital parameters and the MOST secondary periods.  In this paper  $\phi$=0.0 refers to the binary orbital period and denotes the time when the secondary is in
front of the primary (deeper optical minimum) and $\phi$= 0.5 denotes
the time when the primary is approximately in front of the secondary
(shallower optical minimum).  While primary minimum is a definition, the
secondary minimum at $\phi$=0.5 is only approximate, since the orbit is
slightly elliptical and also varies slowly with apsidal motion. The
actual  secondary minimum is currently $\phi$=0.45 (Paper III).  Actual
current quadrature phases are $\phi$=0.23 and 0.78.  To avoid confusion,
we use the phases in this paper that would be assumed with  a circular
orbit (i.e., $\phi$=0 for inferior conjunction, $\phi$=0.5 for superior
conjunction, $\phi$=0.25 and 0.75 for quadrature). Also, we use the
ephemeris of  \citet{mayer10}.   No evidence of
X-ray emission from the tertiary star was seen in any of the \chandra\
observations (Paper I), so the spectra represent only \delOri\ Aa1 and
Aa2.

\section{OVERALL VARIABILITY OF X-RAY FLUX}
\label{section:lc}

The lightcurve of the dispersed \chandra\ spectrum of \delOri\ Aa,
shown in lower part of Fig.\ \ref{most_time}  shows the spectrum was variable throughout in X-ray flux with a maximum
amplitude of about $\pm$15\% in a single observation. During the first
and second observations the X-ray flux varied by $\approx$5-10\%,
followed by a $\approx$15\% decrease in the third and a $\approx$15\%
increase in the fourth.   Note that
none of the X-ray minima in the residual lightcurve aligns with an
optical eclipse of the \delOri\ Aa system.
Fig.\ \ref{most_time} suggests an increase in
overall X-ray flux with time. From the beginning to end of the 9-day campaign
there is a $\approx$25\% increase in the mean count rate. The best linear fit to the entire
lightcurve is 96.04 counts/d  + 0.002 counts/d ∗ HJD, with equal weights
for all points.

\input{period.tex}

We first did a rough period search on the delta Ori Aa Chandra lightcurves using the software package Period04 \citep{lenz05}, and found peaks around 4.8d and 2.1d. This method uses a speeded-up Deeming algorithm \citep{kurtz85} that is not appropriate for sparse datasets such as ours because it assumes the independence of sine and cosine terms valid only for regular lightcurves. Therefore, to verify this preliminary conclusion and get final results, we rather rely on several methods specifically suitable to such sparse lightcurves: the Fourier period search optimally adapted to sparse datasets \citep[][ see Fig. 5]{heck85,gosset01}, as well as variance and entropy methods \citep[e.g.][]{schwarz89,cincotta99}. The results of all these methods were consistent within the errors of each other.

\begin{figure*}[!htb]
\includegraphics[width=2.\columnwidth]{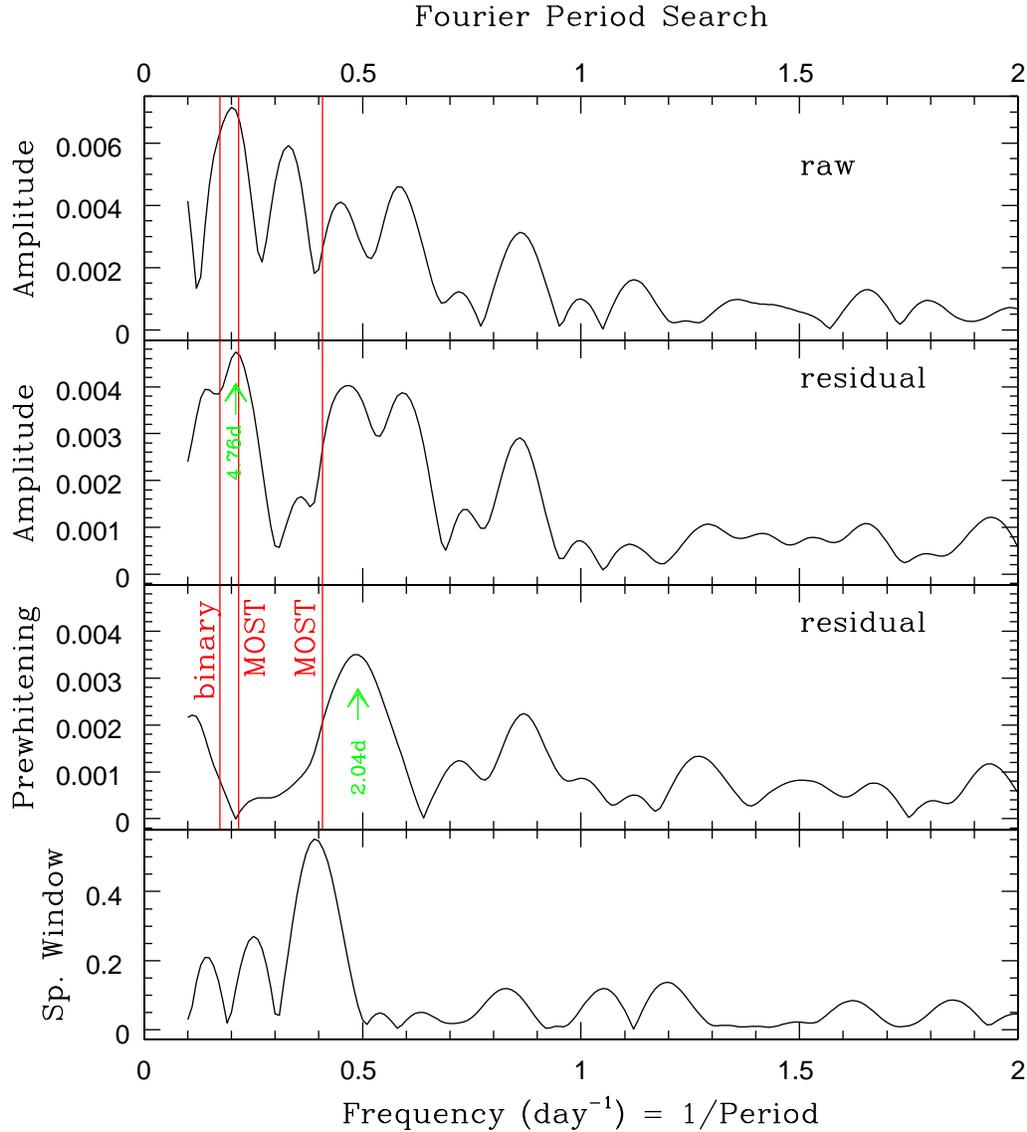}

\caption{Periodograms derived using a Fourier period search adapted for sparse datasets. Frequencies
identified in other wavelengths are shown as red vertical lines.  The
left red vertical line corresponds to the binary period (5.73d), and the
two  strongest secondary MOST periods (4.613d and 2.448d) are indicated by center
and right red vertical lines. {\it Top two panels:} Periodogram for the
raw and residual data, respectively. {\it 3rd panel:} Periodogram for
the residual data after ``cleaning" (prewhitening) of the strongest
signal (4.76d), leaving clearly a second period (2.04d).   {\it Bottom:}
Associated spectral window, showing the relative positions where aliases
may rise.}
\label{fig:period}
 \end{figure*}

\begin{figure}[!htb]
\includegraphics[angle=90,width=1.\columnwidth]{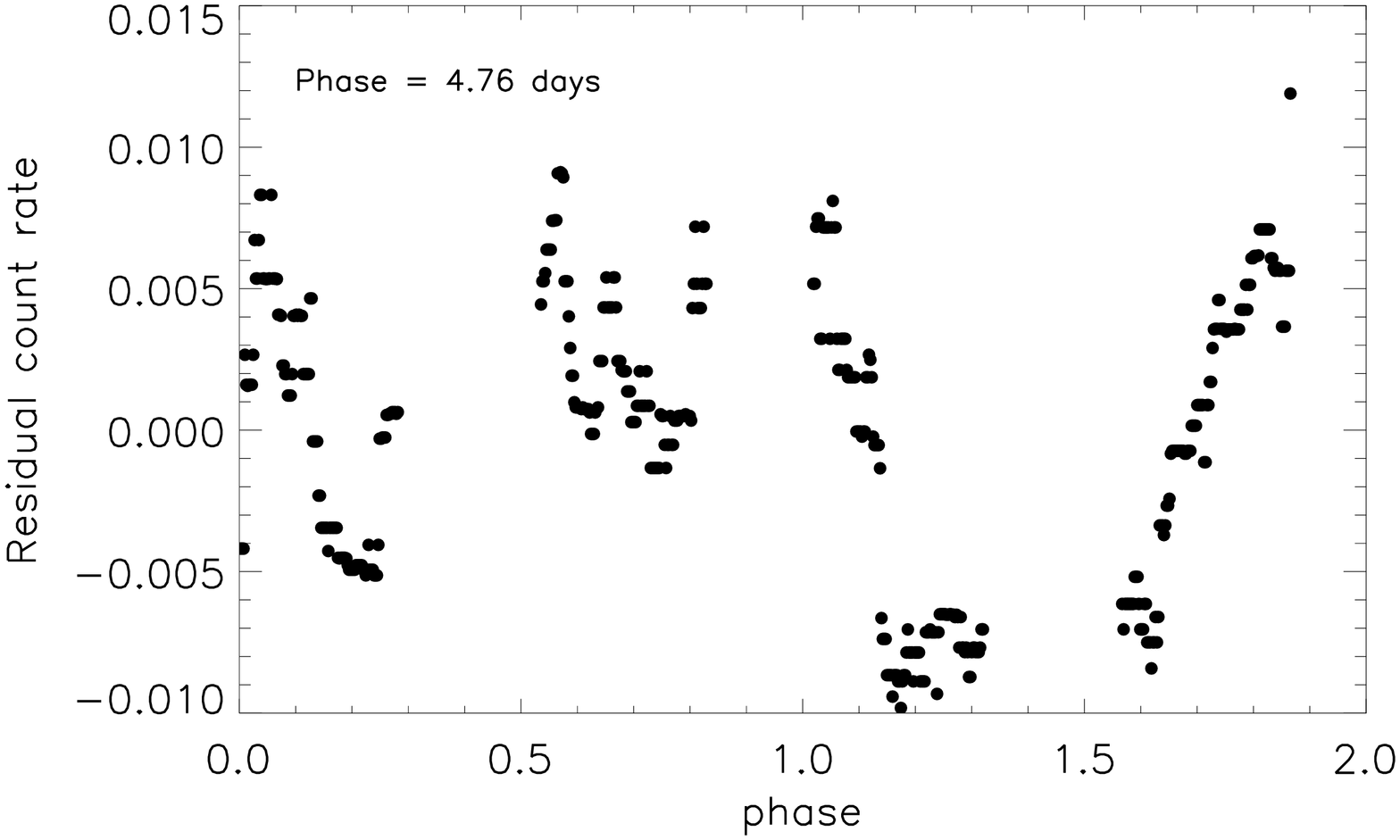}
\caption{The residual count rate lightcurve of 1~ks binned data after
correction for linear fit and filtering by a running median
over $\approx$1/4d bin size.  The x-axis is phase for the 4.76d period found using period search techniques.  See text for explanation.} \label{fig:residual_lc}
\end{figure}

 Using these tools, we first looked for periods in the raw $\cts$ lightcurve data.  A
period of 5.0$\pm$0.3d was found with an amplitude of $7.1\times
10^{-3}$.  We then removed the linear trend described above from the raw data, producing a residual lightcurve. The period searches were repeated, with an identified period of 4.76$\pm$0.3d
(amplitude=$4.7\times 10^{-3}$).  After pre-whitening the residual data in Fourier space for this period, an additional significant period of  2.04$\pm$0.05d (amplitude=$3.5\times
10^{-3}$) was found.  Each of these periods has a
Significance Level  (SL) of $SL<$1\% with the  definition of SL as a
test of the probability of rejecting the null hypothesis given that it
is true.  If the $SL$ is a very small number, the null hypothesis can be
rejected because the observed pattern has a low probability of occurring
by chance.  Fig.\ \ref{fig:period} shows the results of the Fourier
period search method for the raw, residual, and pre-whitened lightcurves, which produced periods of 5.0d, 4.76d, and 2.04d, respectively. Table \ref{table:period} lists the frequency,
period, and amplitude of the  periods.     Fig.\
\ref{fig:residual_lc} shows the residual lightcurve with the period 4.76d plotted on the x-axis.  The residual data were smoothed with a median
filter for the plot only, in order to see the short-term variability
more clearly; the analysis used the unsmoothed residual data
points.

The 5.0d period in the raw data and the 4.76d period in the residual data with the linear trend removed are considered to be the same period because the errors overlap.  Comparing the periods identified in the \chandra\ \del\ Ori Aa data with the MOST optical periods,  the strongest \chandra\ period of 4.76d is consistent within the errors to the MOST$_{F2}$ period of 4.614d (see Table \ref{table:params}).  The \chandra\ period of 2.04d is consistent with the less significant MOST$_{F10}$ period of 2.138d.  There is no evidence of an X-ray period matching the binary period of 5.73d.

Finally, we  again searched for periods including the lightcurve from the original \chandra\
observation of \delOri\ Aa (\obs\ 639 taken in 2001). The lightcurve for obsid 639 alone did not
yield any statistically significant periods  since it covers a much
smaller time interval (60 ks, hence 0.7 d). However, when combined with
the 2013 data, we find similar results as mentioned above for the raw
data, with a period of 5.0d (and its harmonics at 2.5d) providing the
strongest peak. Residual lightcurves were not analyzed with the early
observation because the linear function (which is probably not truly
linear) could not be determined across an 11-year gap.

\section{TIME- AND BINARY PHASE-RESOLVED VARIABILITY}
\label{section:data_analysis}

\subsection{Time-sliced spectra}
\label{section:time_sliced}

The discrete photon-counting characteristic of the ACIS detector allows
the creation of shorter time segments of data from longer observations.
Time-resolved and phase-resolved variability of flux and emission line
characteristics were investigated using a set of short-exposure spectra,
contiguous in time (``time-sliced spectra"), covering the entire
exposure time of the observations, along with individual instrumental
responses to account for any detailed changes in local response with
time, such as might be introduced by the $\approx$1 ks aspect dither
pointing of the telescope.      This was accomplished by reprocessing
the set of  time-sliced data using TGCat software, taking care to align
the zeroth order images among the time-sliced spectra  prior to the
spectral extraction to produce correct energy assignments for the
events. The resulting time segments are of similar exposure time,
$\approx$40 ks, making them easily comparable.  Table \ref{table:slice}
lists the 12 time-sliced spectra, along with the beginning and end time
and binary phase range.   Our \chandra\ observations cover parts of
orbits 396-398 based on the ephemeris.  The integer portion of the
binary phase is in reference to the epoch of the ephemeris used.   The
decimal portion is the binary phase for the specific orbit.

\input{slice_table.tex}

In addition to the 12 time-sliced spectra described above, we  produced
time-sliced spectra of approximately 10 ks in length from the  \chandra\
observations, using the same technique described.  Forty-eight
time-sliced spectra with approximately 10 ks exposure time each were
used in the composite spectral line analysis in Sect.\
\ref{section:composite}.  All other analyses used the 40 ks time-sliced
spectra. We have not included the 2001 \chandra\ observation, obsid 639,
in the following time-resolved emission-line analyses, primarily because
the long-term trend of flux variability seen in our 9-day observing
campaign would make interpretation of this early observation
questionable with respect to flux level.

We describe below several different analyses of the variability of the
dispersed spectral data.    All comparisons in this Section are made to the binary orbital period, not to the periods found in the X-ray flux in Sect. \ref{section:lc}, because we are interested in relating any variability to the known physical parameters of the system and possible effects of the secondary on the emission from the primary wind. First,  statistical tests were performed on
narrow wavelength-binned data for the 12 time-sliced spectra of 40 ks
each to test for variability using $\chi^2$ tests.  We then formally fit
the emission lines in each of these individual 40 ks time-sliced spectra
using Gaussian profiles (Sect.\ \ref{section:fit}), determining fluxes,
line widths, and 1$\sigma$ confidence limits. Subsequent composite line
spectral analysis used the combined H-like ion profiles, as well as
\ironsevt, to evaluate the flux, velocity, and line width as a function
of binary phase (Sect.\ \ref{section:composite}).  Finally, we looked
for variability in the $fir$-inferred radius ($R_{fir}$) of each ion, as
well as X-ray temperatures derived from the H-like to He-like line
ratios (Sect.\ \ref{section:fi}).

\subsection{Narrow-band  fluxes and variability} \label{section:bin}

\input{tbl_fprops.tex}

For the following statistical analysis, we used the narrow
wavelength-binned bands in the 12 time-sliced, 40 ks spectra described
above.  The count rates for a standard set of narrow wavelength bins
were output from TGCat processing. The parameters of the bins between
2.5 \ang\ and 22.20 \ang\ are listed in Table \ref{table:fprops}. We
searched for variations using a series of $\chi^2$ tests, trying several
hypotheses (constancy, linear trend, parabolic trend) and checked the
improvement when more free parameters were used.  The Significance Level
($SL$) is defined in Sect.\ \ref{section:lc}.  Five bands were
significantly variable  when compared to a constant value, i.e. $SL \le$1\%: (1) the continuum centered at
4.9 \ang, (2) \sfir, (3) \sifir, (4) \irontwenty\ (10.4-12 \ang), and
(5) \nefir. A further 4 bands are marginally variable, i.e.
$1\%<SL<5$\%: (1) \mgH, (2) the continuum centered at 8.8 \ang, (3)
\neH, and (4) the continuum centered at 14.925 \ang. When compared to a linear trend, all but \irontwenty\ were significantly variable, and when compared to a quadratic trend, all but \nefir, \irontwenty, and \sifir\ were significantly variable, although in all cases $SL<5$\%.

As an additional test, we directly compared the spectra one to another.
Using a $\chi^2$ test on the strongest wavelength bins, with spectra
binned at 0.02 \ang, variability was significant for lines (or in the
regions of lines): \sifir, \mgH, \mgfir, \nefir, and the zone from
10.4-12 \ang\ (corresponding to \irontwenty).

\clearpage
\begin{figure*}[!htb]
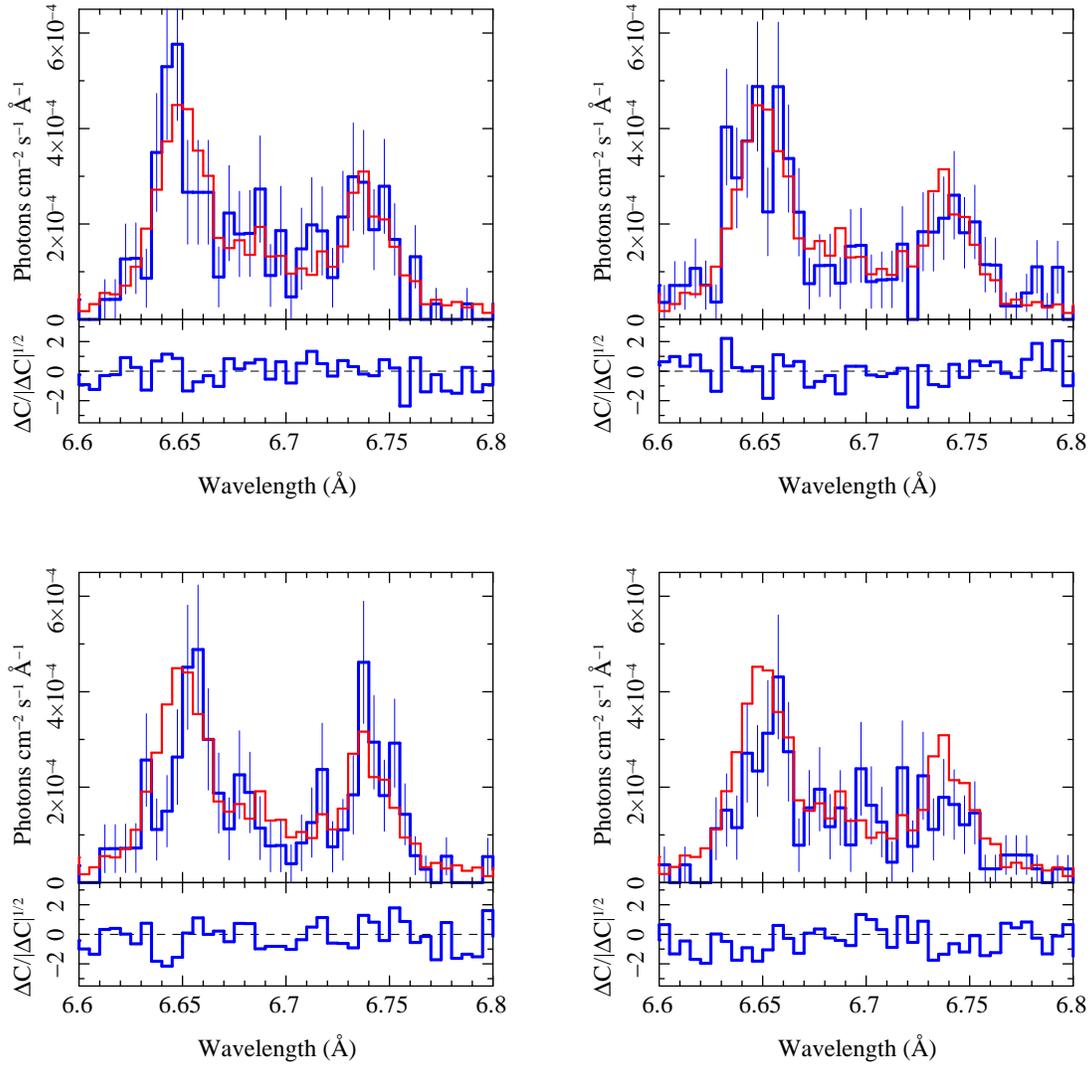

\centering
\includegraphics[width=\columnwidth]{dori_si13_vs_mean_heg_meg-02.ps_page_7}
\includegraphics[width=\columnwidth]{dori_si13_vs_mean_heg_meg-02.ps_page_9} \\
\includegraphics[width=\columnwidth]{dori_si13_vs_mean_heg_meg-02.ps_page_11}
\includegraphics[width=\columnwidth]{dori_si13_vs_mean_heg_meg-02.ps_page_1} \\
\caption {\sifir\ profile (blue) overplotted with the mean \sifir\ profile (red) of all
time-sliced spectra.  Upper panel left: Phase is centered at 0.091 and
is 0.15 wide; Upper panel right: Phase is centered at 0.254 and is 0.15
wide; Lower panel left: Phase is centered at 0.576 and is 0.15 wide;
Lower panel right: Phase is centered at 0.644 and is 0.15
wide.}\label{fig:profiles}
\end{figure*}

In summary,  \sfir, \sifir, \nefir, and \irontwenty\ were variable in
both of the above tests.  An example of the \sifir\ lines for several
time slices is shown in Fig.\ \ref{fig:profiles}, demonstrating the
observed variability.  As noted later, \nefir\ is contaminated by Fe
lines, which may contribute to the variability.  A few other emission
lines as well as some continuum bandpasses may also be variable, but
with lower confidence. As an additional confirmation of variability for
one feature, \sifir, we performed a two-sample Kolmogorov Smirnov (KS)
test \citep{press02} on each time slice against the complementary
dataset (to ensure the datasets are independent). With the criteria that
an emission line is variable if the null hypothesis is $\le$ 0.1, and
that a line is not variable if the null hypothesis is $\ge$ 0.9, the
only spectrum where the K-S test suggested real variability (at about
2\% probability of being from the same distribution) was \sifir\ at
$\phi$=0.11.  Note that the KS test shows that there is variability, but
not what is varying, such as flux or centroid.

\subsection{Fitting of Emission Lines} \label{section:fit}

For each of the 12 time-sliced spectra of 40 ks each, the H- and He-like
lines of S, Si, Mg, Ne, and O, as well as \ironsevt\ 15.014 \ang, were
fit using the Interactive Spectral Interpretation System
\citep[ISIS;][]{houck00}.  Only Gaussian profiles were considered
because  (1) Gaussian profiles are generally appropriate at this
resolution for thin winds at the signal-to-noise level of the
time-sliced spectra (with some exceptions, noted below), and (2)
previous studies indicated Gaussian profiles provided adequate fits to
the emission lines for both the early \chandra\ observation of \delOri\
Aa \citep{miller02} and for the combined spectrum from 2012 (Paper I).
We note that lines in the spectrum of the combined HETGS data showed
some deviations from a Gaussian profile (Paper I).

The continuum for each time-sliced spectrum was fit by using the same
3-temperature APEC model derived in Paper I.  This model allowed for
line-broadening and a Doppler shift.    Some abundances were fit in
order to minimize the residuals in strong lines.    An N$_H$ of
$0.15\times 10^{22}\cmmtwo$ (Paper I) was fixed for the value of the
total absorption.   Only the continuum component of this model was used
for continuum modeling in the following analysis.

Fits were determined by folding the Gaussian line profiles through the
instrumental broadening using the RMF and ARF response functions, which
were calculated individually for each time-sliced spectrum. All first
order MEG and HEG counts, on both the plus and minus arms of the
dispersed spectra, were  fit simultaneously.  For most H-like ions, the
line centroid, width, and flux were allowed to vary.  In a few cases
where the S/N was low, the line center and/or the width was fixed to
obtain a reasonably reliable fit based on the Cash statistic.  For the
He-like line triplets, the component lines were fit simultaneously with
the line centroid of the recombination ($r$) line allowed to vary, and
the centroids for the intercombination ($i$) and forbidden ($f$) lines
forced to be offset from the $r$ line centroid by the theoretically
predicted wavelength difference.  The individual flux and width values
of the triplet components for the He-like ions were allowed to vary
except for a few cases of low S/N when the width value of the $i$ line
and $f$ line were forced to match that of the $r$ line to obtain a
reasonable statistical fit. The reduced Cash statistic using subplex
minimization was used to evaluate each fit.  For most emission lines
with good signal-to-noise, the reduced Cash statistic was 0.95-1.05.  A
few of the lines with poor signal-to-noise had a reduced Cash statistic
as low as 0.4.  Confidence limits were calculated at the 68\% (i.e.
$\pm1\sigma$) level for each parameter of each line, presuming that
parameter was not fixed in the fit.

In most cases, the line profiles in the time-sliced spectra were well
fit with a simple Gaussian.  In a few cases, a profile might be better
described as flat-topped with steep wings.    Broad, flat-topped lines
are expected to occur when the formation region is located relatively
far from the stellar photosphere, where the terminal velocity has almost
been reached.  In such cases, Gaussian profiles are expected to fit
rather poorly. Occasionally a second Gaussian profile was included for a
line if a credible fit required it.     If more than one Gaussian
profile was used to fit a line, the total flux recorded in the data
tables is the sum of the individual fluxes of the Gaussian components
with the errors propagated in quadrature.

\input{flux_table_land.tex}

For the case of \nefir\ where  lines from \ironsevt\ and \ironninet\ provided
significant contribution in the wavelength region of the fit,  these additional
Fe lines were fit separately from the \nefir\ component lines.  The Fe lines fit in this region  were  \ironninet\ at 13.518 \AA, \ironninet\ at 13.497 \AA, and \ironsevt\ at 13.391 \AA, with  \ironninet\ at 13.507 \AA\ and \ironninet\ at 13.462 \AA\ included at their theoretical intensity ratios to \ironninet\ at 13.518 \AA.
Also, the \neH\ line is blended with
an \ironsevt\ line.  We have assumed this \ironsevt\ component
contributes flux to the \neH\ line equal to 13\% of the \ironsevt\ at
15.014 \ang\ \citep{walborn09}.  A final correction was applied to the
\sifir-$f$ line because the Mg Lyman series converges in this wavelength
region.  Using theoretical relative line strengths, we assumed the
\sifir-$f$ line flux was overestimated by 10\% of the measured flux of
the \mgH\ L$\alpha$ line.

\begin{figure*}[!htb]
\centering
\includegraphics[angle=90,width=2.5in]{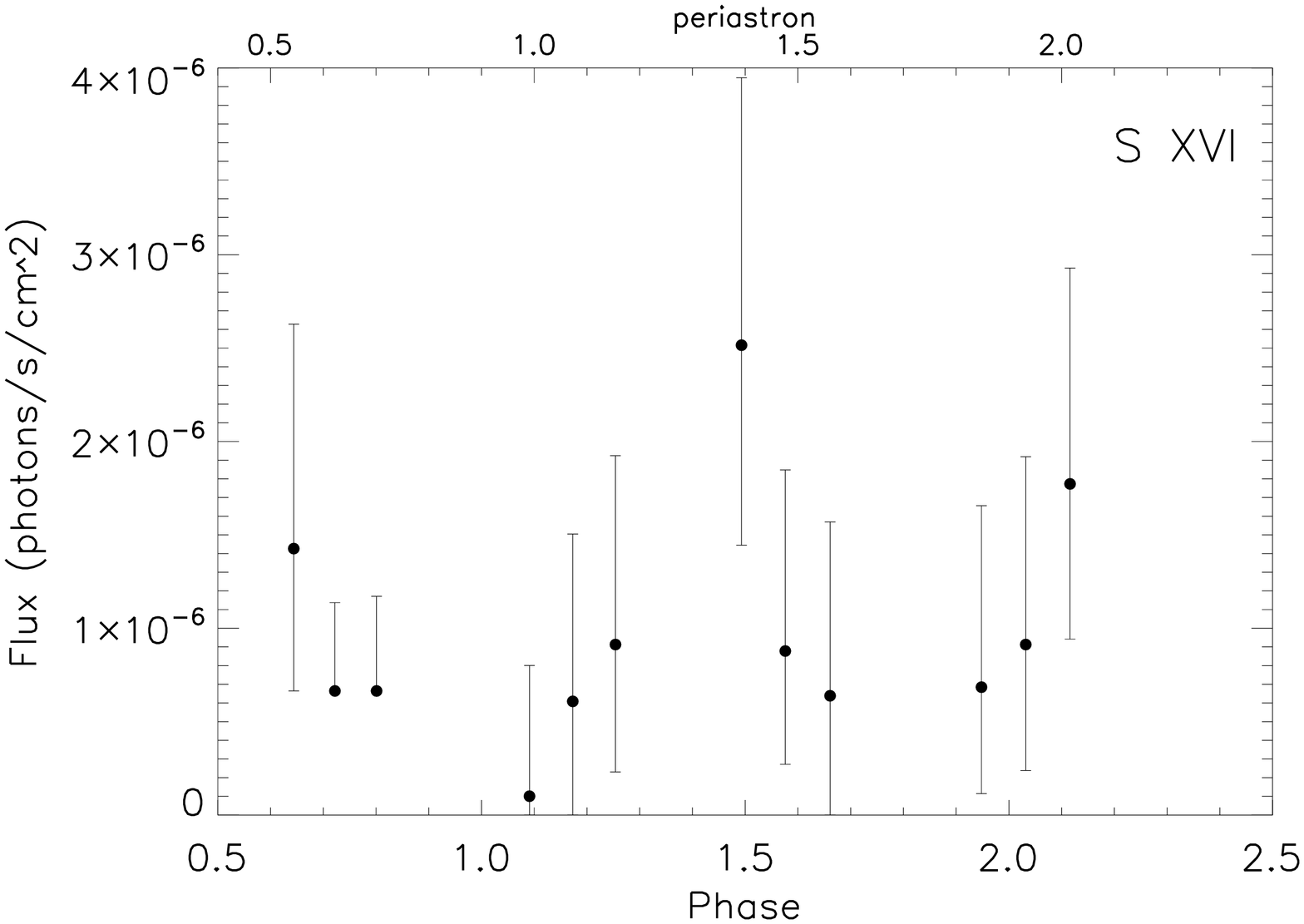}
\hfill
\includegraphics[angle=90,width=2.5in]{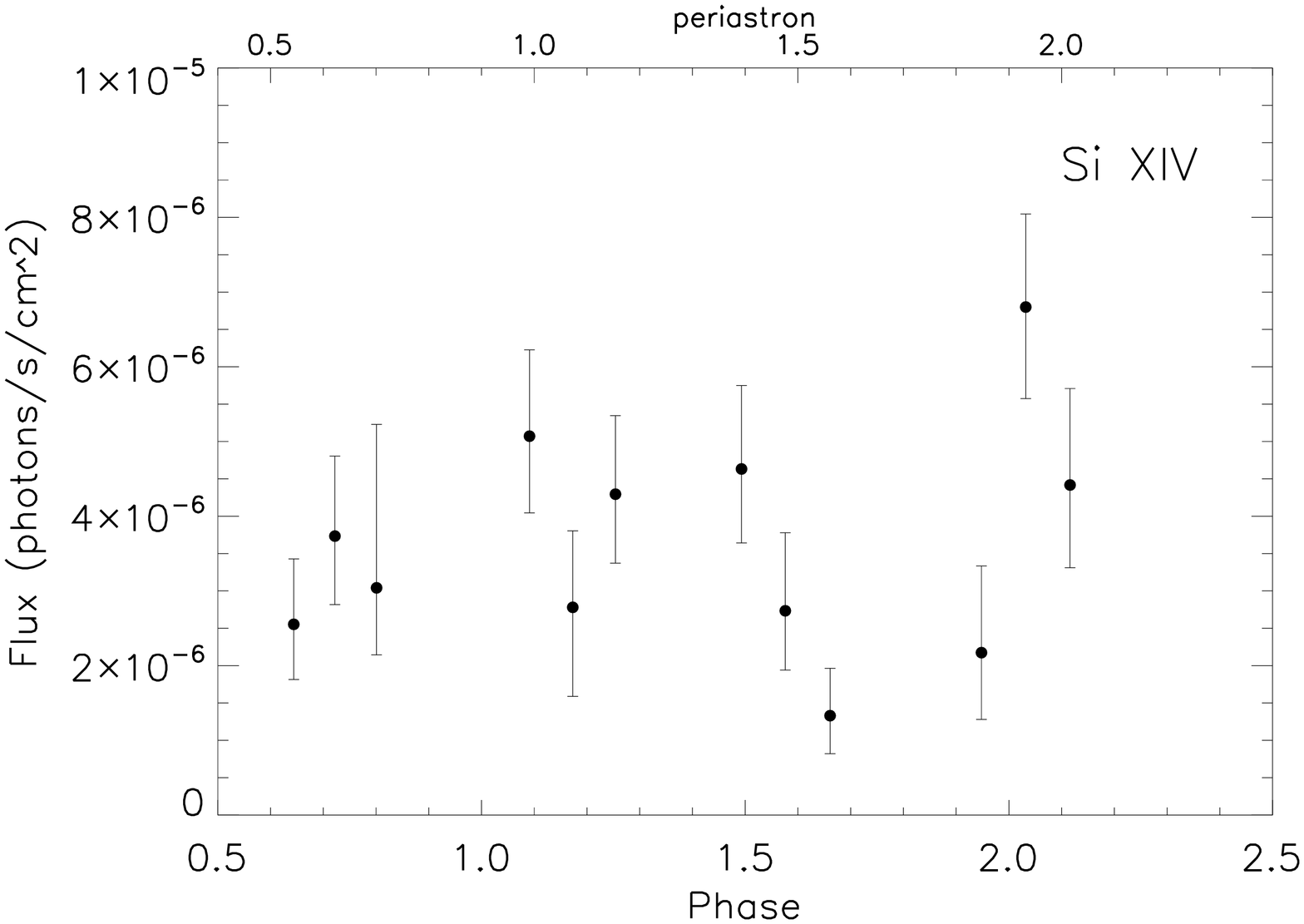} \\
\includegraphics[angle=90,width=2.5in]{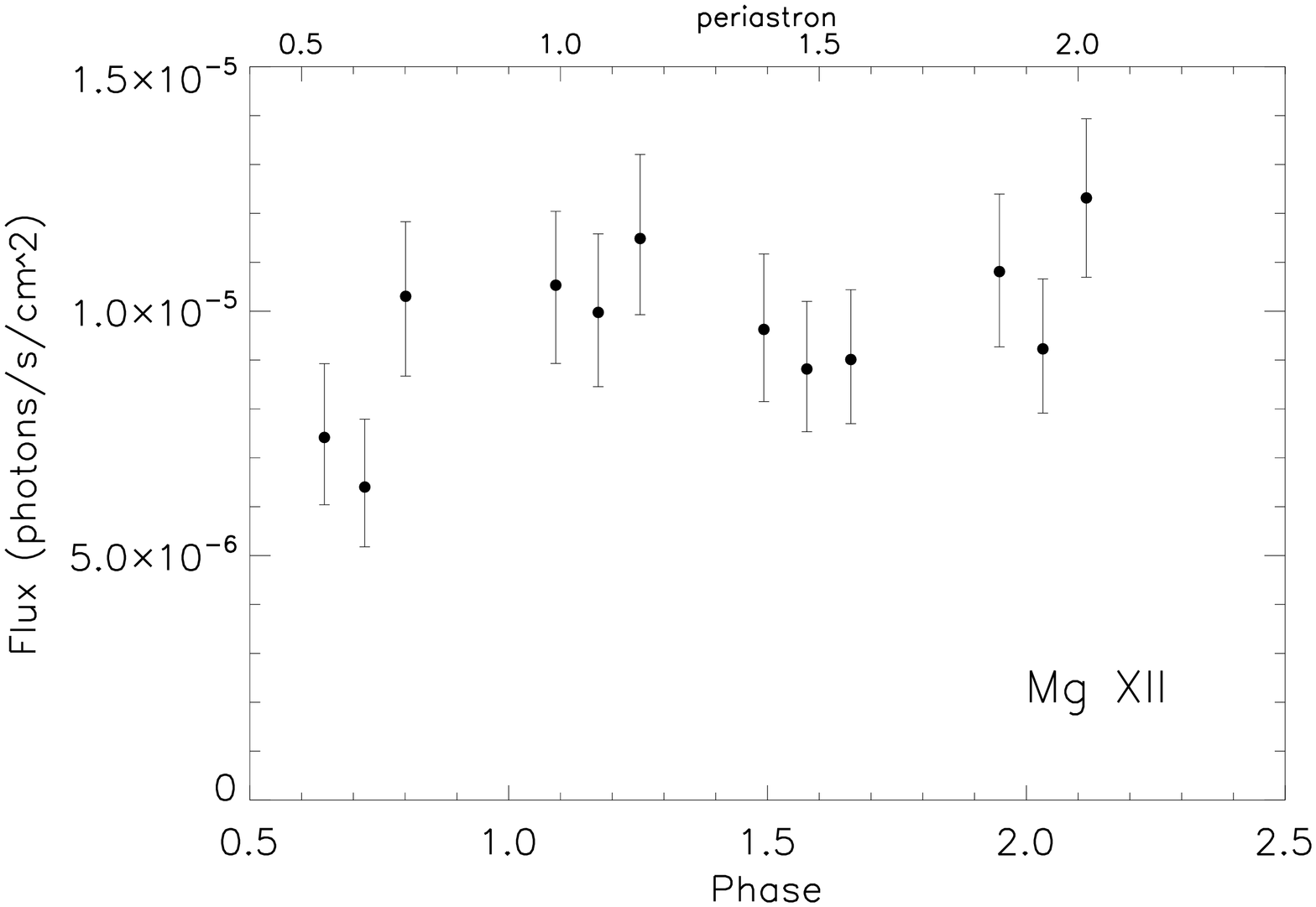}
\hfill
\includegraphics[angle=90,width=2.5in]{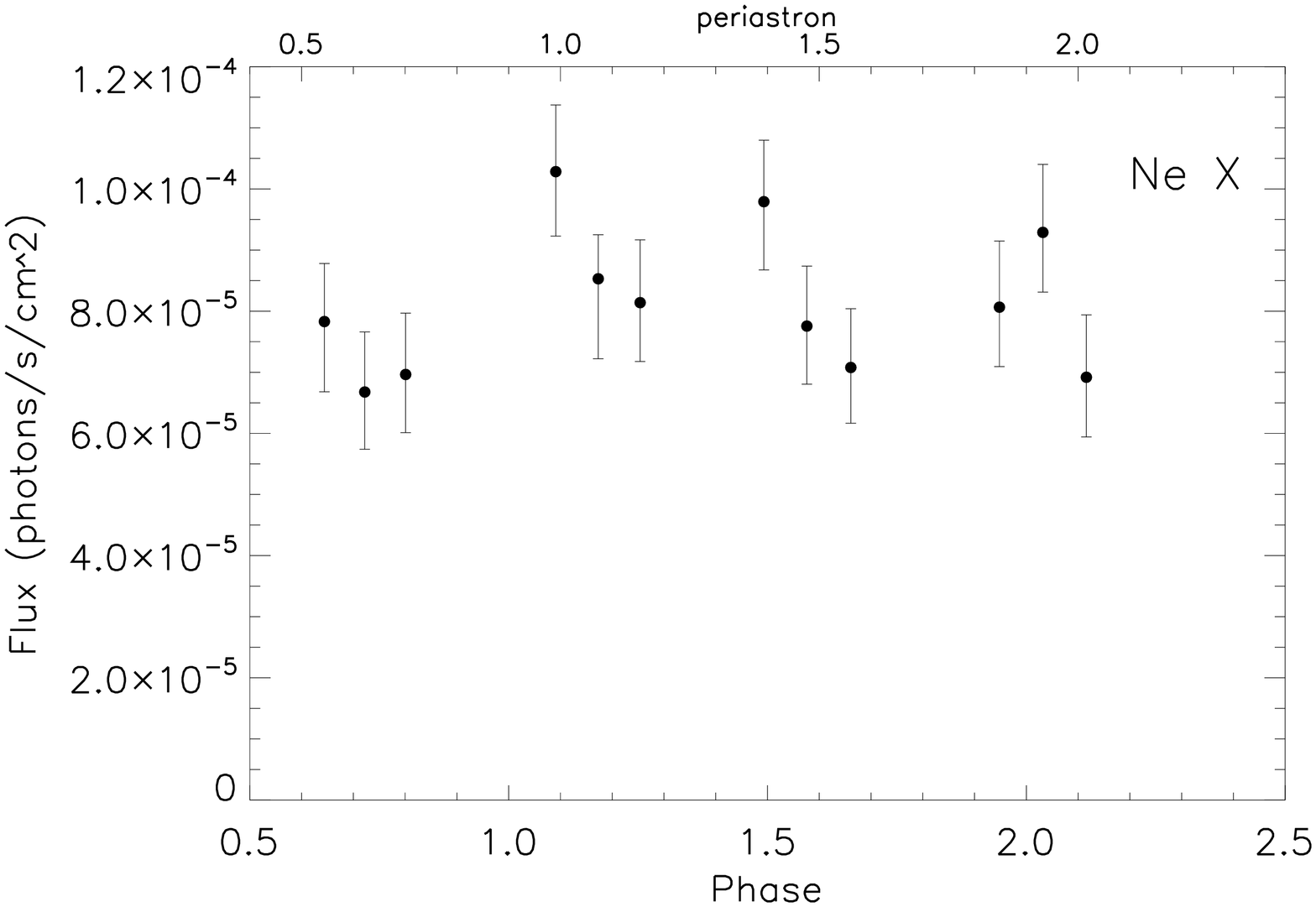} \\
\includegraphics[angle=90,width=2.5in]{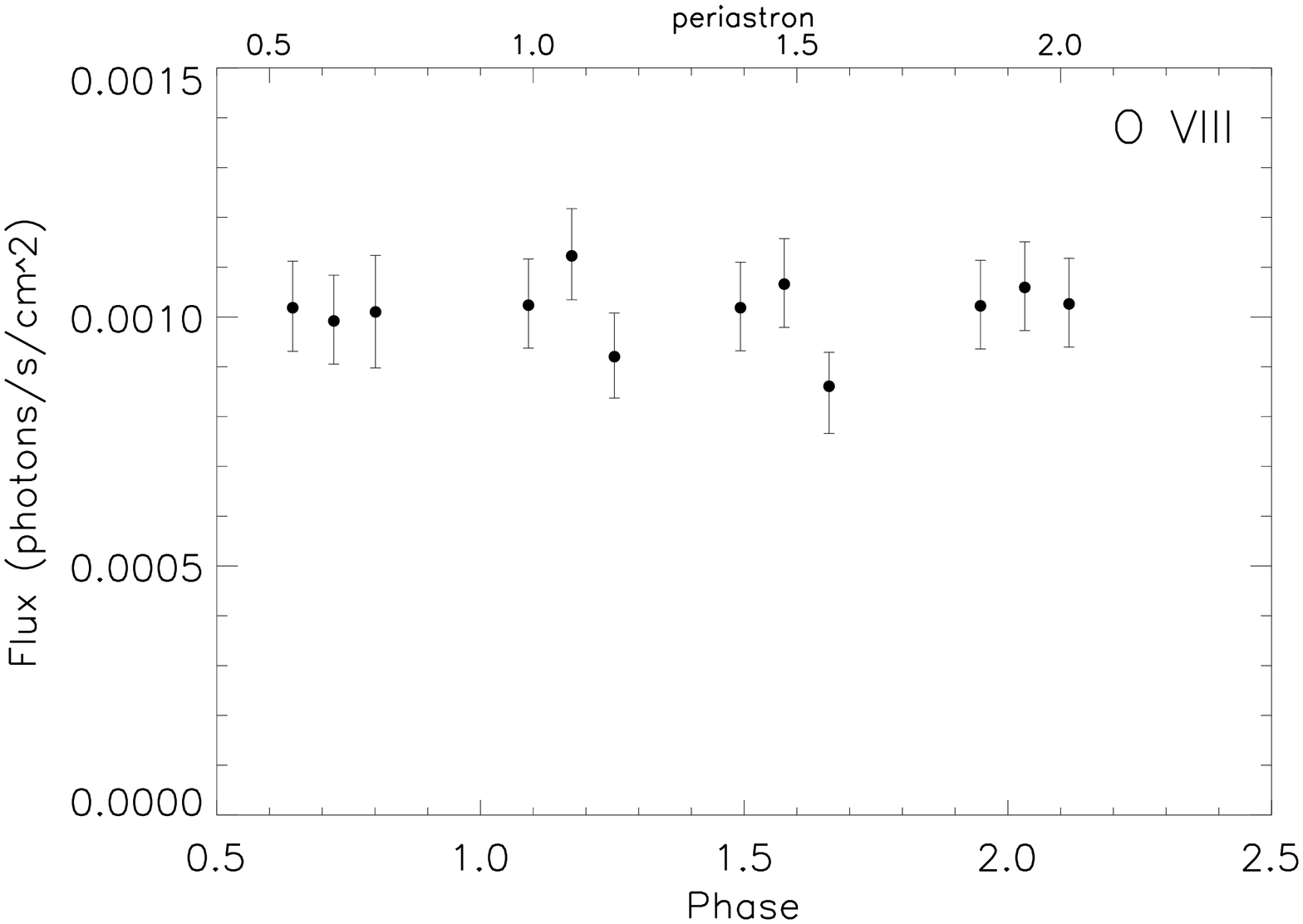} \\

\caption{Fluxes of H-like emission lines based on Gaussian fits vs phase.
Errors are 1$\sigma$ confidence limits.   Phase with respect
to periastron is indicated at the top of the plot. }
\label{flux1}
\end{figure*}

\begin{figure*}[!htb]
\centering
\includegraphics[angle=90,width=2.5in]{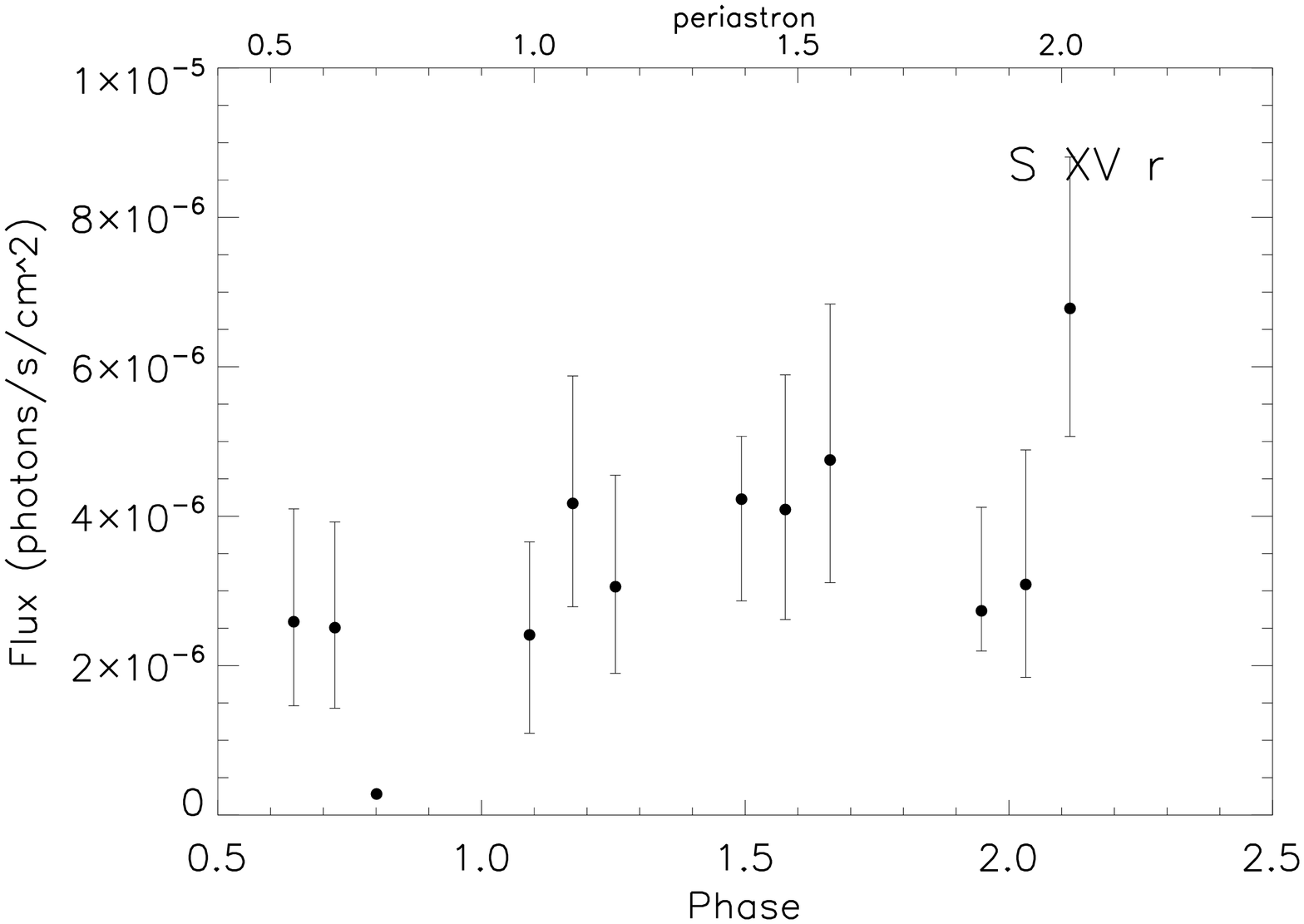}
\hfill
\includegraphics[angle=90,width=2.5in]{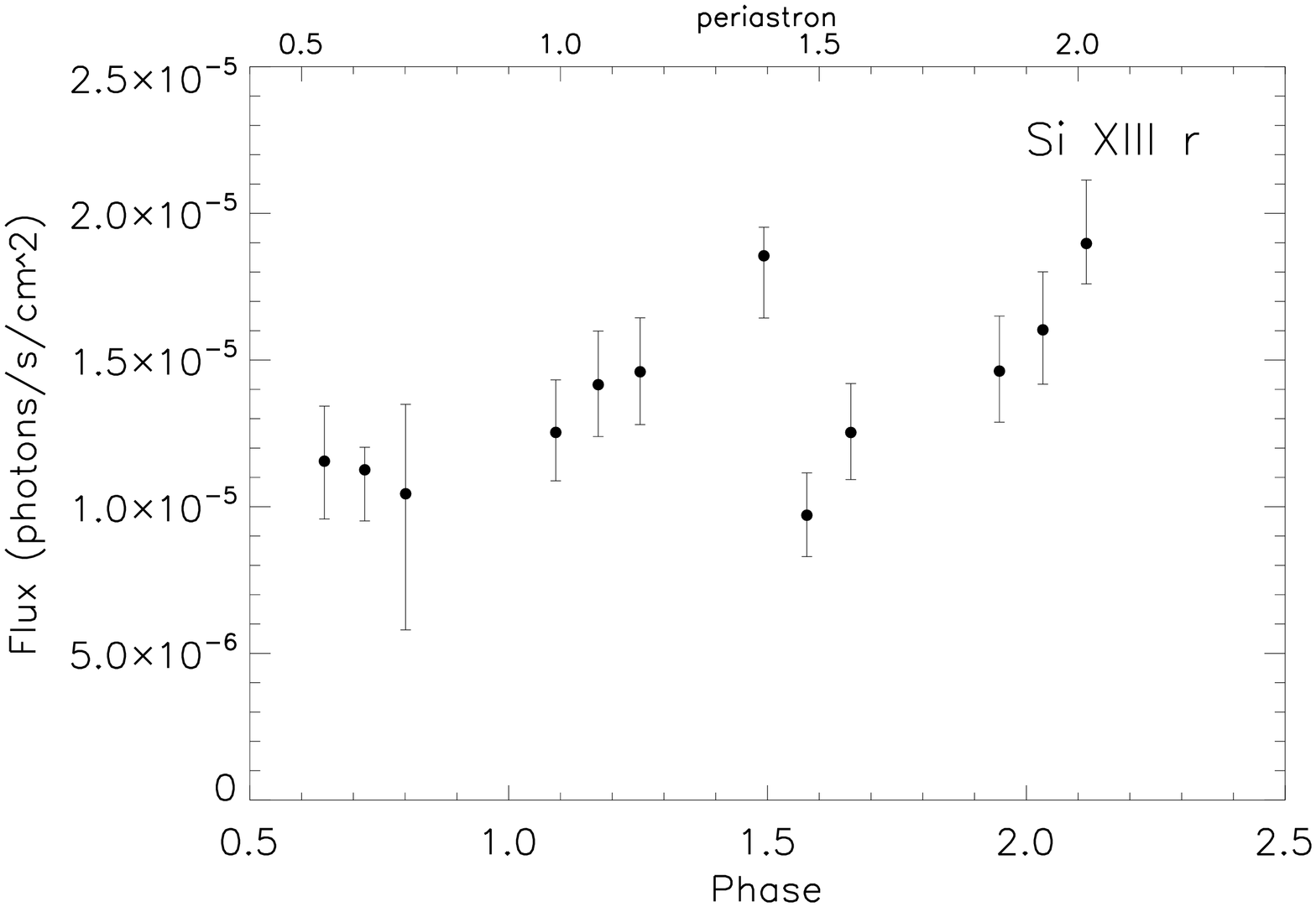} \\
\includegraphics[angle=90,width=2.5in]{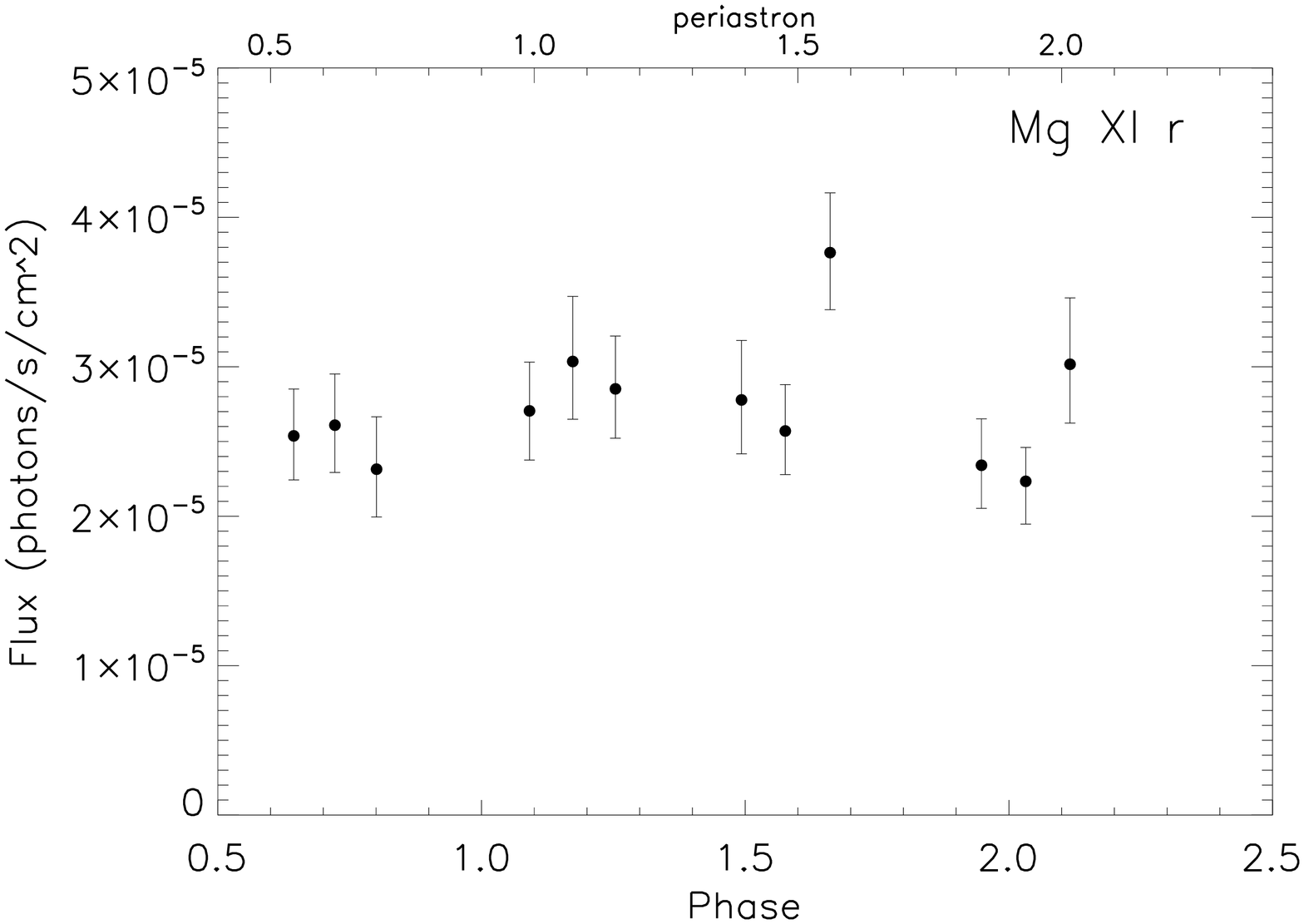}
\hfill
\includegraphics[angle=90,width=2.5in]{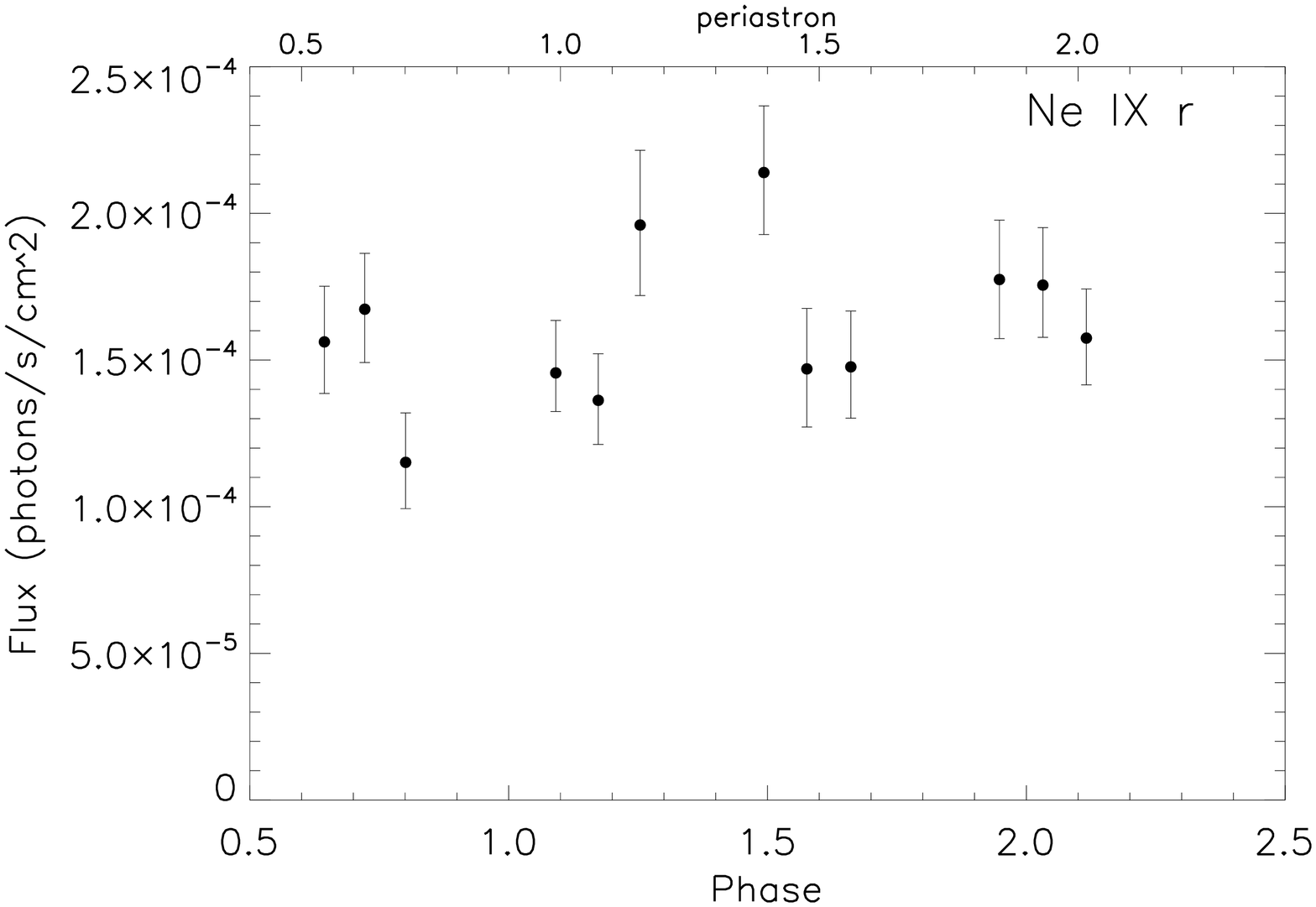} \\
\includegraphics[angle=90,width=2.5in]{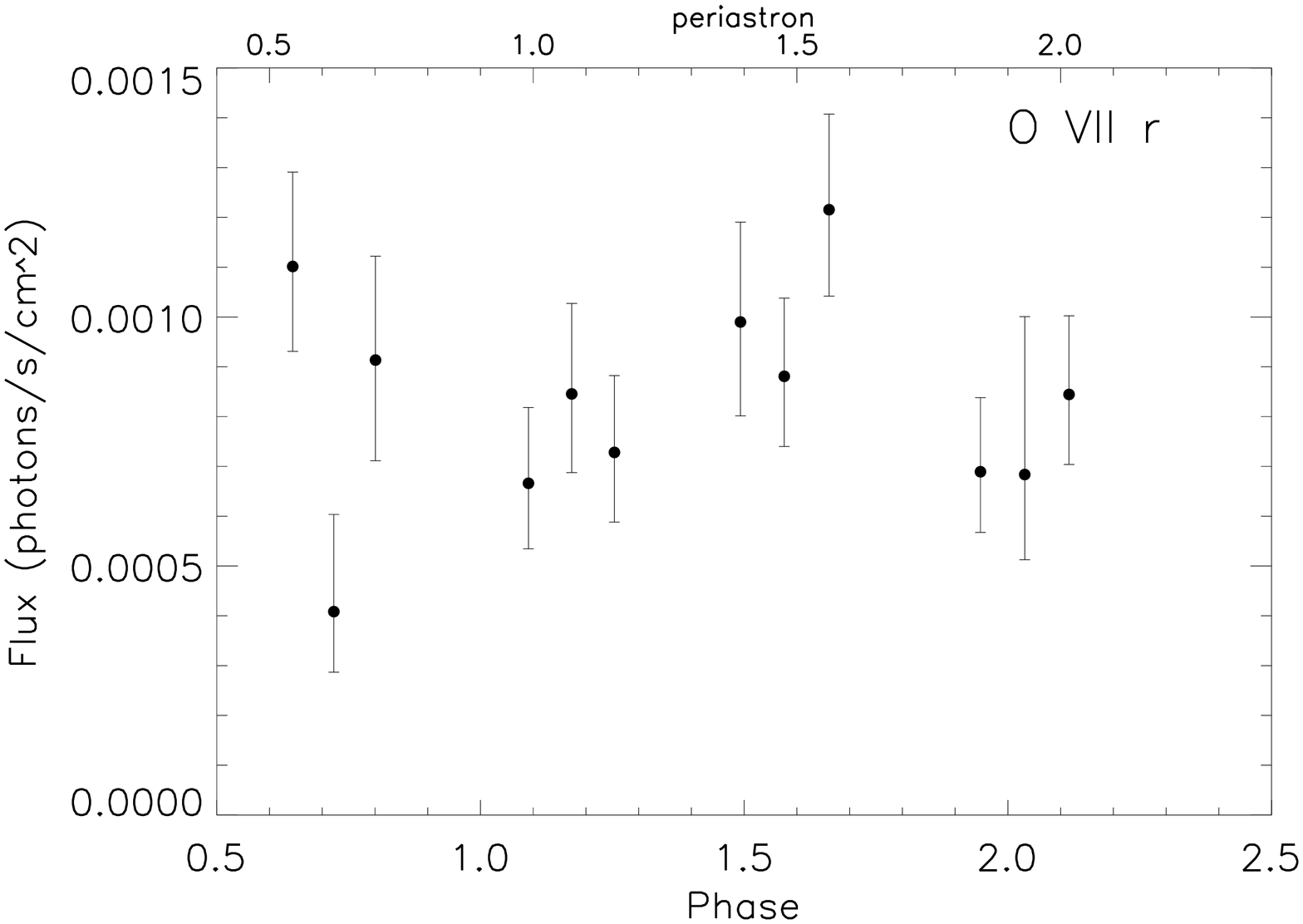} \\

\caption{Fluxes of He-like $r$ emission lines based on Gaussian fits vs
phase.  Errors are 1$\sigma$ confidence limits.}\label{flux2}
\end{figure*}

The flux values and confidence limits are tabulated in Table
\ref{table:flux:s} for S and Si lines,  Table \ref{table:flux:mg} for Mg
and Ne lines, and Table \ref{table:flux:o} for O  and \ironsevt\ lines.
To summarize the results, Fig.\ \ref{flux1} shows a comparison of the
fluxes of the H-like ions.  The error bars for \sH\ are
quite large.  \siH\ shows a peak at about $\phi$=0.0 and a somewhat
lower value at about $\phi$=0.6.  \mgH, \neH, and \oH\ are essentially
constant.  Fig.\ \ref{flux2} shows the fluxes for the He-like
$r$ lines.     \sfir-$r$ has a maximum at about $\phi$=0.1, with lower
flux in the range $\phi$=0.5-0.8.  The flux values for \sifir-$r$ are
larger for $\phi$=0.0-0.4 than for the range $\phi$=0.5-0.8.  \ofir-$r$
shows an apparent increase in flux in the $\phi$=0.5-0.7 range.
\mgfir-$r$ and \nefir-$r$ are relatively constant.  \nefir\ was
consistently variable in the narrow-band statistical tests.  In this
Gaussian fitting of the lines we have fit and removed the contaminating
Fe lines in \nefir, possibly removing the source of variability in this
triplet.  We note that the points in Fig.\ \ref{flux1} and Fig.\
\ref{flux2} are from three different orbits of the binary.  The increase
in flux with time discussed in Sect.\ \ref{section:lc} has not been
taken into account in these fitted line fluxes, either in the plots or
the data tables, so care must be taken in their interpretation.

\subsection{Spectral Template and Composite Line Profile Fitting}
\label{section:composite}

\begin{figure}[!htb]
 \centering
 \includegraphics[width=0.95\columnwidth]{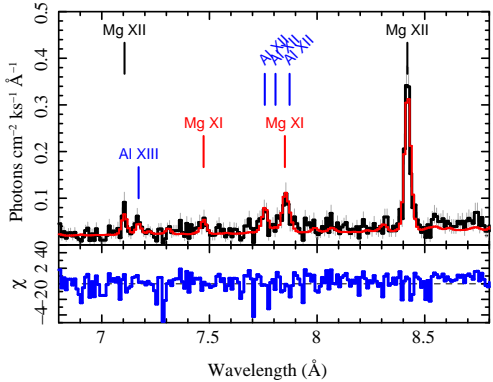}
 \caption{A portion of the HEG spectrum for the entire exposure,
 after fitting an APEC template to the $8.3$--$8.6$\,\ang\  region.  The temperatures, relative normalizations, and absorption
 were frozen parameters.  The Doppler shift, line width, and
 normalization were free.  We show the resulting model evaluated over
 a bit broader region that was fit to demonstrate the applicability
 of the model to the local spectral region.  Other regions will not
 necessarily be well represented by the same parameters.
  \label{fig:apectemplate}}
\end{figure}

In order to improve the signal-to-noise ratio in line fits in the
time-sliced data, we used two methods to fit multiple lines
simultaneously. In the first method, we adopted a multi-thermal APEC
\citep{Smith:01, Foster:Smith:Brickhouse:2012} plasma model which
describes the mean spectrum fairly well (see Table~\ref{tbl:apecmodel}),
and used this as a spectral template, allowing the fits to the Doppler
shift, line width, and overall normalization to vary freely. This is a simpler model than the more
  physically based APEC model defined in Paper I
  since here it need not fit the spectrum
  globally, but is only required to fit a small region about a
  feature of interest.  To demonstrate this,
  Figure~\ref{fig:apectemplate} shows an example after fitting {\em
  only} the $8.3$--$8.6$\,\AA\  region for the \mgH\
  feature's centroid, width, and normalization for the entire
  exposure.  The temperatures were not varied, and the relative
  normalizations of the three components were kept fixed, as was the
  absorption column.  We can see that this provides a good local
  characterization of the spectrum, and so will be appropriate for
  studying variation of these free parameters in local regions as a
  function of time or phase.  For any such fit, other regions will
  not not necessarily be well described by this model.

\begin{figure}[!htb]

\includegraphics[width=0.9\columnwidth]{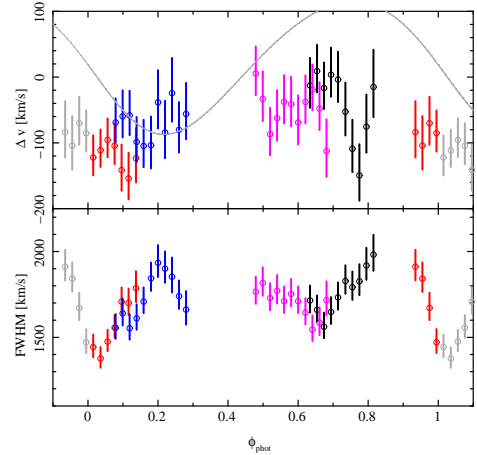}
\caption{The mean emission line Doppler velocity (top: points with error
bars), primary radial velocity (top, sinusoidal curve), and mean line
width for \neH\ (bottom) derived by fitting spectra in phase bins with
an APEC template, allowing the Doppler shift, line width, and
normalization to vary freely.  Data from the individual \chandra\
observations are differentiated with colors.  Error bars (1$\sigma$) are
correlated over several bins since a running average was used over 3-10
ks bins.} \label{fig:template}
\end{figure}

These fits
used spectra extracted in 10 ks time bins (about $0.02$ in phase), but
were fit using a running average of three time bins. We primarily used
the H-like Lyman$\,\alpha$ lines, as well as some other strong and
relatively isolated features (see Table~\ref{tbl:linesfit}). The results
for the interesting parameters, the mean Doppler shift of the lines and their
widths, are shown in Fig.\ \ref{fig:template}. The smooth curve in the
top panel is the binary orbital radial velocity curve.  We clearly see a
dip in the centroid near $\phi= 0.8$, as well as significant changes in
average line width.

\begin{deluxetable}{rr}
\tabletypesize{\scriptsize}
\tablecolumns{2}
\tablewidth{0pt}
\tablecaption{Plasma model  parameters  used for spectral  
template fitting.}\label{tbl:apecmodel}
\tablehead{
\multicolumn{2}{c}{Temperature Components\tablenotemark{a}} \\
\colhead{$T$} & \colhead{$Norm$} \\
}

\startdata

2.2& 8.16 \\
6.6& 1.90 \\
19.5& 0.226 \\
    \hline
  
    \multicolumn{2}{c}{Relative Abundances\tablenotemark{b}}\\

    Elem.& $A$ \\
    \hline

    Ne& 1.2\\
    Mg& 0.7\\
    Si& 1.6\\
    Fe& 0.9\\
    \hline
    \multicolumn{2}{c}{Total Absorption\tablenotemark{c}}\\
    \hline
    $N_H$& 0.15\\

\enddata

 \tablenotetext{a}{Temperatures are given in MK, and the normalization
 is  related  to the volume emission measure, $VEM$, and distance,
 $d$, via  $VEM = 10^{14}$ $(4\pi d^2) \times \Sigma_i Norm_i$}.
 \tablenotetext{b}{We give elemental abundances relative to the solar
 values  of  \citet{Anders:89} for those significantly different from
 1.0.  (These are not rigorously determined abundances, but related
 to  discrete temperatures adopted  and actual abundances.)}
 \tablenotetext{c}{The total absorption is given in units of $10^{22}\cmmtwo$.}

 \end{deluxetable}

 \begin{deluxetable}{rl}

\tabletypesize{\scriptsize}
\tablecolumns{2}
\tablewidth{0pc}
\tablecaption{Lines used in CLP  analysis\tablenotemark{a}}
\label{tbl:linesfit}
\tablehead{
\colhead{$\lambda_0$}    &  \colhead{Feature}  \\
 \colhead{ \ang} & \colhead{} \\
}
\startdata

    6.182&  Si~XIV \\
    8.421&  Mg~XII \\
    10.239&  Ne~X \\
    11.540&  Fe~XVIII \\
    12.132& Ne~X \\
    14.208&  Fe~XVIII  \\
    15.014&  Fe~XVII \\
    15.261& Fe~XVII \\
    16.005&  Fe~XVIII$+$O~VIII \\
    16.780&  Fe~XVII \\
    17.051&  Fe~XVII \\
    18.968&  O~VIII \\

 \enddata
 \tablenotetext{a}{Lines used in
    the ensemble fitting with Composite Line Profile or spectral
    template methods.  Both HEG and MEG were used at wavelengths shorter
    than 16 \ang.  The region widths were 0.20 \ang, centered on each
    feature.}
    \end{deluxetable}

\begin{figure}[!htb]
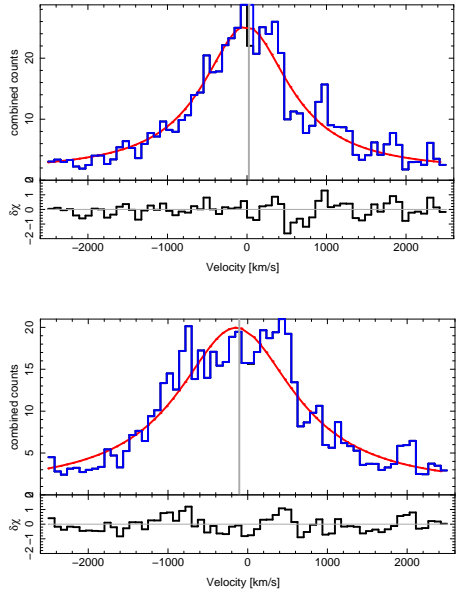

\includegraphics*[width=0.9\columnwidth, viewport=0 0 480
290,angle=0]{plt_r1-03_phi068.ps}
\includegraphics*[width=0.94\columnwidth, viewport=0 0 500
290,angle=0]{plt_r1-03_phi078.ps}
\caption{An example of Composite Line
Profiles for two different phases with different centroids, 0.68 (top)
and 0.78 (bottom), as defined by the photometric ephemeris.  In each
large panel, the histogram is the observed profile, the smooth curve is
the fit.  In the small panel below each are the residuals.}
\label{fig:clplines}
\end{figure}

In the second method, we rebinned regions around selected features to a
common velocity scale and summed them into a Composite Line Profile
(``CLP'').  While this mixes resolutions (the resolving power is
proportional to wavelength and is different for HEG and MEG) and blends
in the CLP, the mix should be constant with phase and be sensitive to
dynamics as long as line ratios themselves do not change. Hence, we can
search for phased variations in the line centroid.  This technique has
been applied fruitfully in characterizing stellar activity in cool stars
\citep{Hoogerwerf:al:2004,Huenemoerder:Testa:al:2006}.  The CLP profiles
were computed in phase bins of $0.01$, but grouped by 5 bins for
fitting, thus forming a running average.  We used the same lines as in
the template fitting. In Fig.\ \ref{fig:clplines} we show an example of
composite line profiles, and fits of a Lorentzian (since the composite
profile is no longer close to Gaussian) plus a polynomial to determine
centroid and width.  This method, while less direct than template
fitting, did confirm the trend seen in line velocity in the template
fitting.

\begin{figure}[!htb]
\centering
\leavevmode
\includegraphics[scale=0.4]{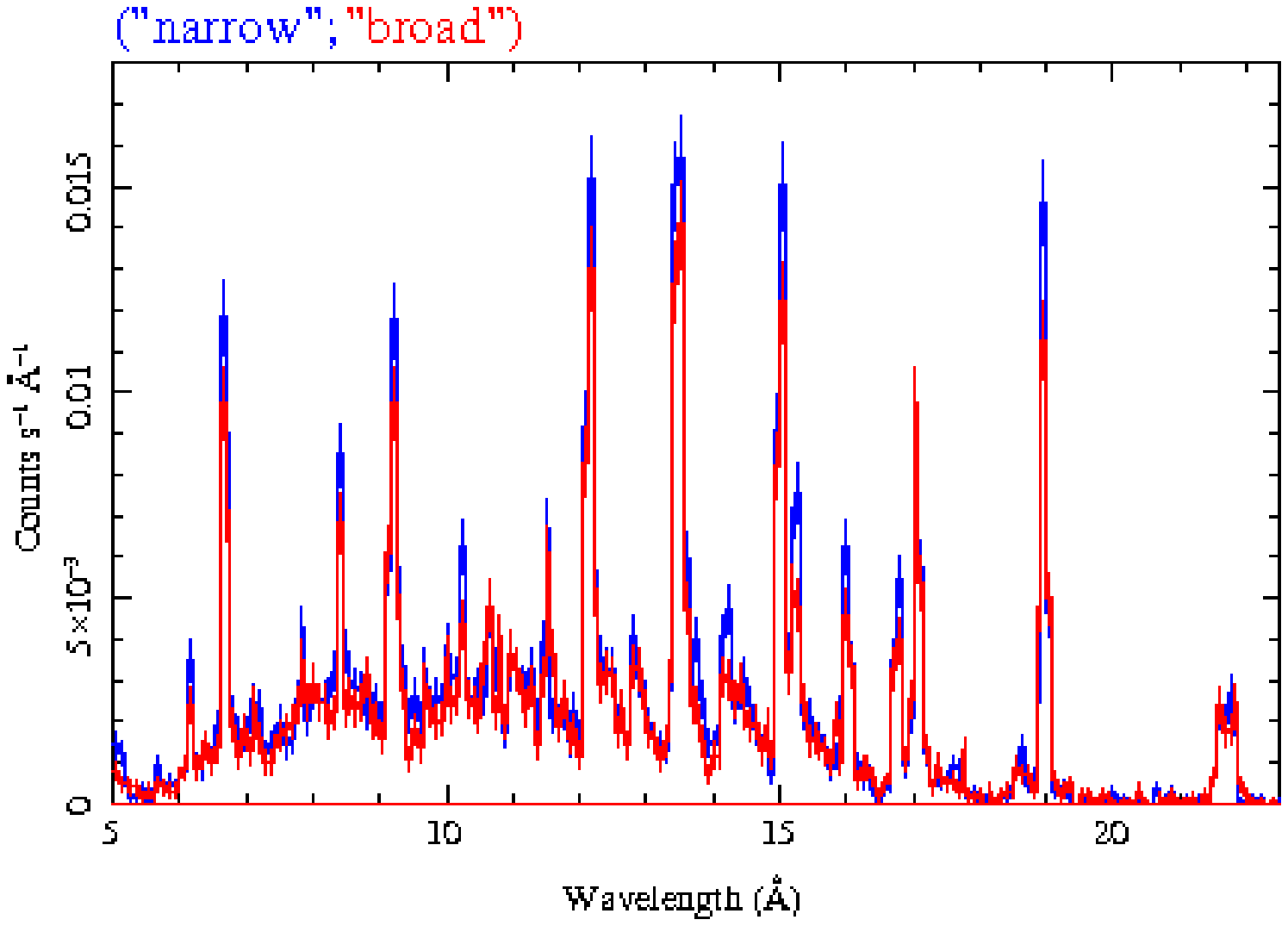} \\
\includegraphics[scale=0.4]{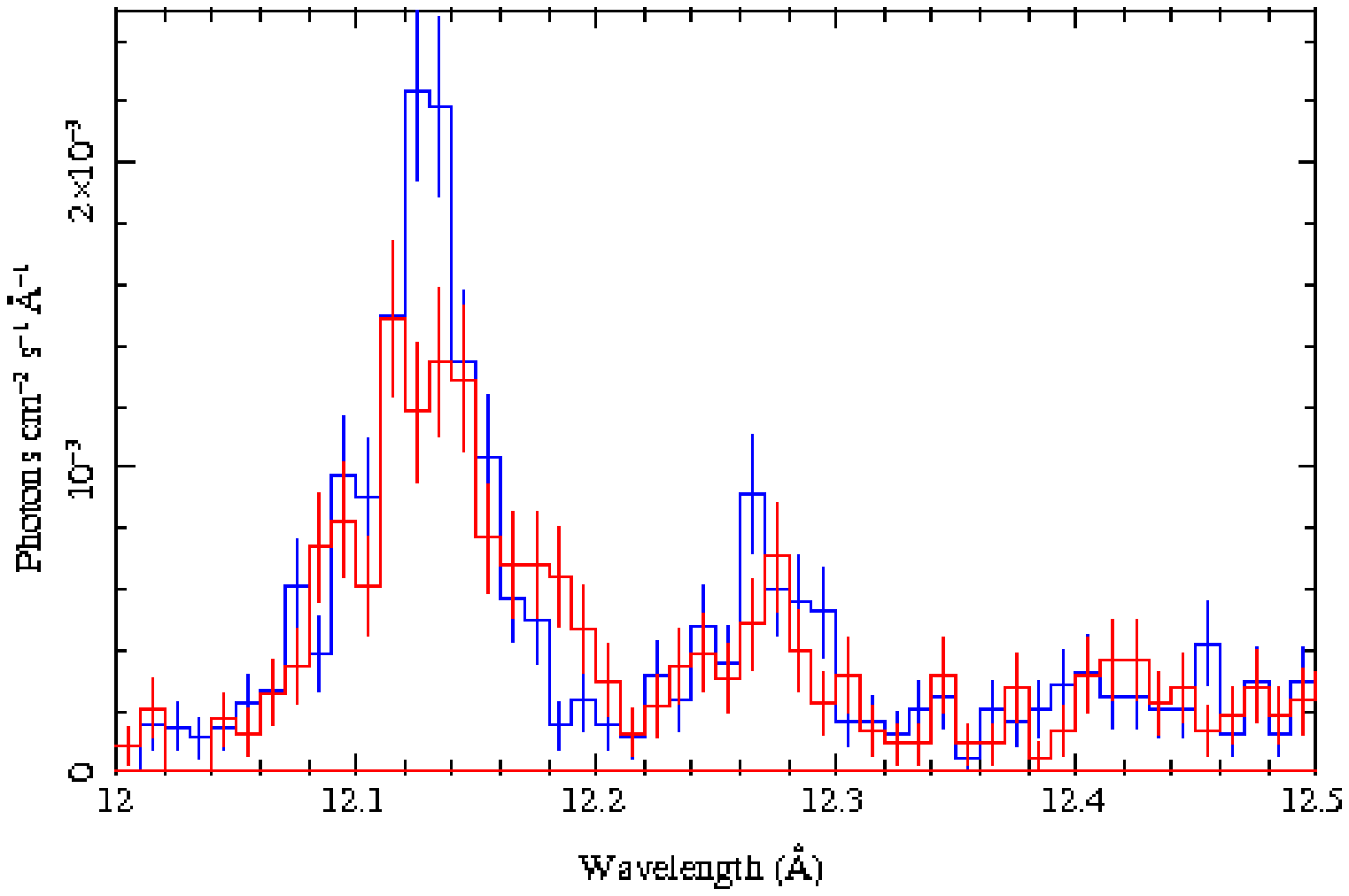} \\
  \caption{Examples of broad and narrow emission lines for selected
  wavelength regions.  Plots are constructed from counts per bin data
  without continuum removal.  For this comparison, the 10 ks time-sliced
  spectra have been combined to represent $\phi$=0.0 (blue) and the
  quadrature phases (red). } \label{fig:linewidth_1}
  \end{figure}

The template-fit line width result is very interesting in that it shows
a significantly narrower profile near $\phi\approx0$ than at other
phases.  Given the trends in width, (low near phase 0.0, high near 0.2
and 0.8) 10 ks time-sliced spectra were grouped in these states and then
compared (Fig.\ \ref{fig:linewidth_1}). The plots show the narrow state
in blue and the broad state in red. The lines are all sharper, except
for  the line at 17 \ang\ (and maybe Si XIV 6 \ang), in the narrow
state. The top panel of Fig.\ \ref{fig:linewidth_1} shows a heavily
binned overview, and the lower panel shows a comparison of the \neH\
line profiles at $\phi$=0 and at quadrature phases.

The line width variability was confirmed by comparing the average
spectrum at phases near $\phi=0.0$ with the average spectrum at other
phases.  The changes were primarily in a reduced strength of the line
core in the phases when the lines are broad, with little or no change in
the wings.

\begin{figure}[!htb]
\centering
\leavevmode
\includegraphics[width=1.0\columnwidth]{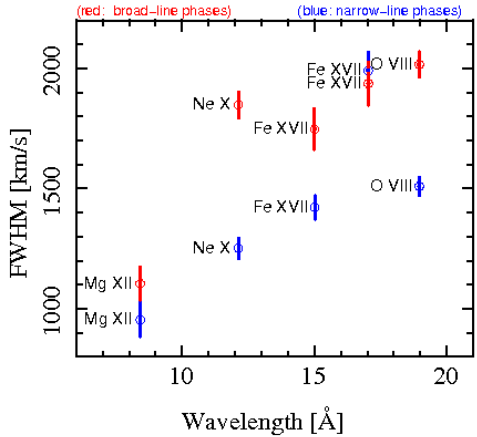}
\caption{Comparison of FWHM in \ks\ for several emission lines in the
time-sliced spectra of \delOri\ Aa.  For this comparison, the 10 ks
time-sliced spectra have been combined to represent $\phi$=0.0 (blue)
and the quadrature phases (red).  Note that these spectra do not have
continuum removal.  Gaussian plus polynomial line fitting was used on
the time slices to determine the line width.} \label{fig:linewidth_2}
\end{figure}

Fig.\ \ref{fig:linewidth_2} shows the trend vs.\ emission line.  Except
for \neH\ and \ironsevt\ 17 \ang, there is a trend for larger
differences in the FWHM with increasing wavelength. Note that
temperature of maximum emissivity goes roughly inversely with
wavelength; wind continuum opacity increases with wavelength.  The
increasing trend is typical of winds, since the opacity causes longer
wavelength lines to be weighted more to the outer part of the wind where
the velocity is higher. Gaussian fit centroids show the lines are all
slightly blueshifted, which could be consistent with skewed wind
profiles.  The ``narrow" group is near the primary star radial velocity
shift. Radial velocities of the lines are all roughly consistent with
-80 \ks, except perhaps \neH, which is blended with an Fe line. The
dependence between line width and binary phase was confirmed
independently by moment analyses of the individual lines, and was also
suggested by the CLP analysis above.
%
%

\subsection{X-ray Emission Line Ratios }\label{section:fi}

The He-like ions provide key plasma diagnostics using the relative
strengths of their $fir$ (forbidden, intercombination, resonance) lines
by defining two ratios\footnote{These lines are often designated as w,
x, y, z in order to highlight the fact that the i-line emission is
produced by two transitions (x+y) such that $R = z/(x + y)$ and $G= (x +
y + z)/w$}: the $R$ ratio =$f/i$ and the $G$ ratio=$(i+f)/r$.
\citet{gabriel69} demonstrated that the $f/i$ and $G$ ratios are
sensitive to the X-ray electron density and temperature, respectively.
These ratios have been used extensively in stellar X-ray studies. In
addition, the presence of a strong UV/EUV radiation field can change the
interpretation of the $f/i$ ratio from a density diagnostic to a
measurement of the radiation field geometric dilution factor, i.e.,
effectively the radial location of the X-ray emission from a central
radiation field \citep{blum72}. The $f/i$ ratio is known to decrease in
the case of a high electron density and/or high radiation flux density,
which will de-populate the upper level of the $f$-line transition
(weakening its emission) while enhancing the $i$-line emission. For hot
star X-ray emission, the $f/i$ ratio is controlled entirely by the
strong UV/EUV photospheric radiation field. The first analysis of an O
supergiant HETG spectrum by \citet{waldron01} verified that the observed
X-ray emission is distributed throughout the stellar wind and
demonstrated that density effects could only become important in high
energy He-like ions if their X-rays are produced extremely close to the
stellar surface. Thus  the $f/i$ ratio can be exploited to determine the
onset radius or $fir$-inferred radius ($R_{fir}$ in units of R$_*$) of a
given ion via the geometric dilution factor of the photospheric
radiation field  \citep{waldron01}. In addition, there are basically two
types of $fir$-inferred radii,  ``localized" (point-like) or
``distributed." The first detailed distributed approach was developed by
\citet{leutenegger06} assuming an X-ray optically thin wind. For a given
observed $f/i$ ratio, the localized approach predicts a larger
$R_{fir}$ as compared to the distributed approach \citep[see discussion
in][]{waldron07}. Since all X-ray emission lines scale as the electron
density squared, all line emissions are primarily dominated by their
densest region of formation. However, in the case of the $fir$ lines, an
enhanced $i$-line emission can only occur deep within the wind (high
density), whereas the majority of the $f$-line emission is produced in
the outer wind regions  at lower densities. The $r$-line emission is
produced throughout the wind.

Another X-ray temperature-sensitive line ratio is the H-like to He-like
line ratio (H/He) as explored in several hot-star studies
\citep[e.g.][]{schulz02,miller02,waldron04,waldron07}. However, a wind
distribution of X-ray sources implies a density dependence (i.e., the
H-like and He-like lines may be forming in different regions) and a
dependence on different wind X-ray absorption effects.  Thus, the
temperatures derived from H/He ratios may be higher than their actual
values  \citep[see][]{waldron04,waldron07}.

Our line ratio analysis is based on the approach given by
\citet{waldron07}. The $f/i$ ratios for each He-like ion in each
time-sliced spectrum are tabulated in Tables \ref{tab:FISI} -
\ref{tab:FIO}.  We did not include S in this analysis because the flux measurement errors are
large and the flux ratio errors are extremely large or unbounded.
We calculated the $fir$-inferred radii ($R_{fir}$) and
H/He-inferred temperatures ($T_{HHe}$) versus phase for the 12
time-sliced spectra, as determined by the Gaussian line fitting. All
radii were determined by the point-like approach and a TLUSTY
photospheric radiation field with parameters T$_{eff}$=29500 kK and log G=3.0.
The model $f/i$ ratios and H/He ratios
used to extract $R_{fir}$ and $T_{HHe}$ information take into account
the possible contamination from other lines. For all derived $R_{fir}$,
we assume that the He-like ion line temperature is at its expected
maximum value.

\begin{figure}[!htb]
\centering
\includegraphics[width=0.9\columnwidth]{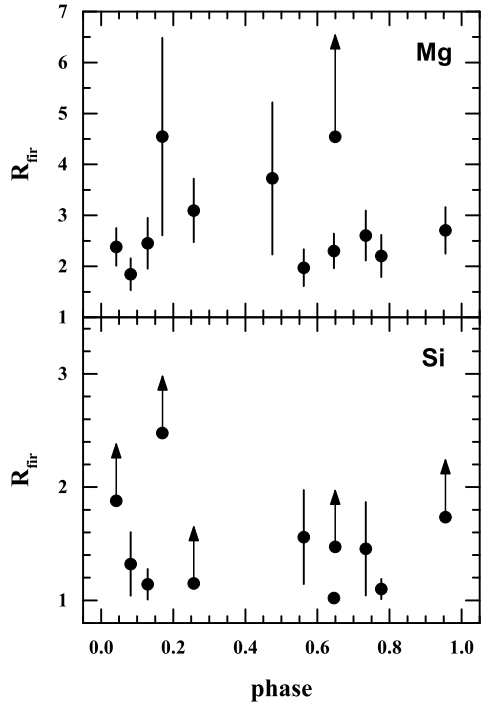}
\caption{Phase dependence of the derived  Mg and Si $fir$-inferred radii
($R_{fir}$) for the 12 time-sliced spectra.  Upper and lower limits are shown as arrows.
See text for
model details.}\label{onemg}
\end{figure}

\begin{figure}[!htb] \centering
\includegraphics[width=0.9\columnwidth]{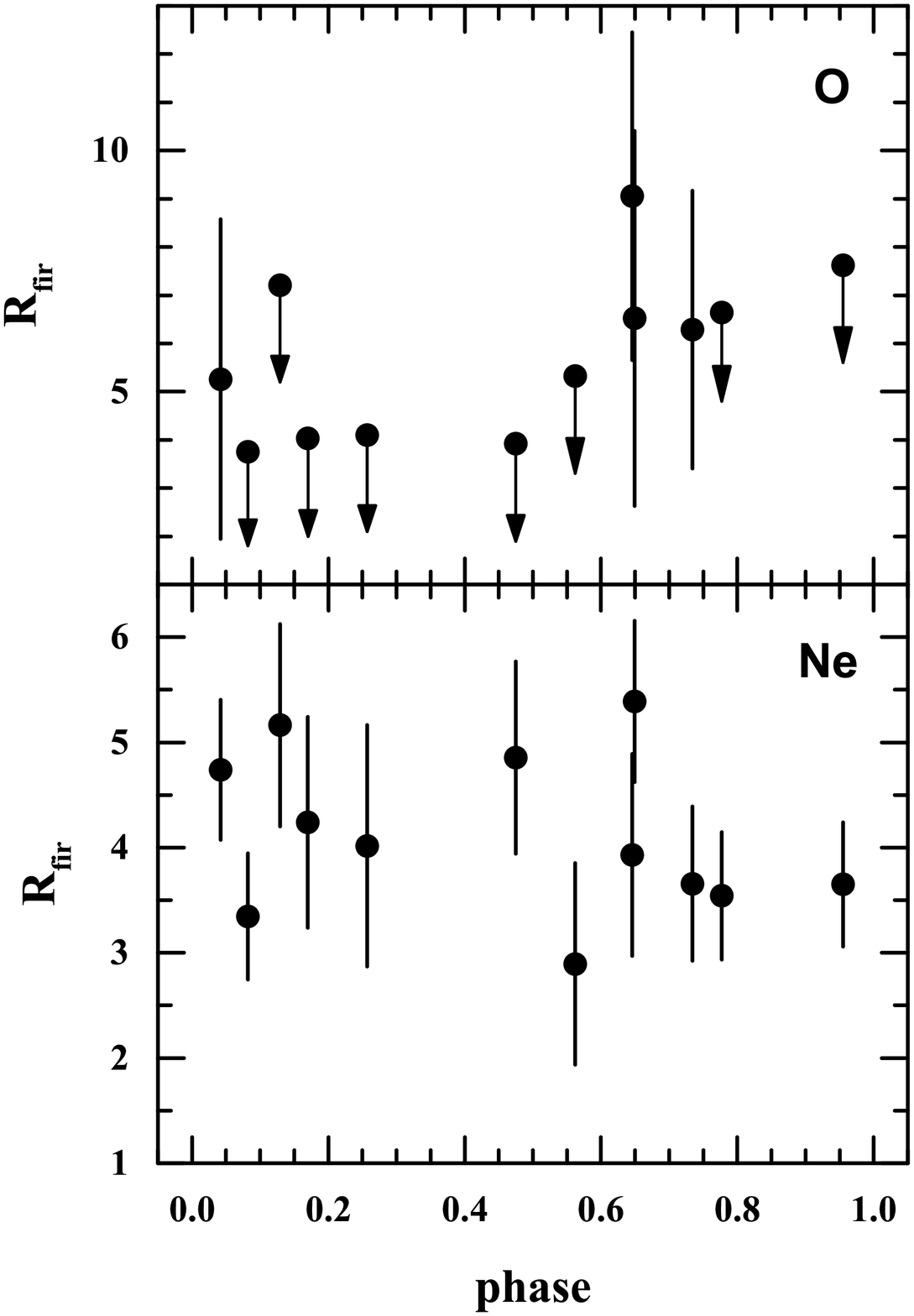}
\caption{Phase dependence of the derived O and Ne $fir$-inferred radii
($R_{fir}$) for the 12 time-sliced spectra.  Upper and lower limits are shown as arrows.
See text for model details.}\label{ssi} \end{figure}

The $fir$-inferred radii for  Mg and Si are plotted in Fig.\
\ref{onemg} and for Ne and O in Fig.\ \ref{ssi}.
In all cases the derived $R_{fir}$ is based on the average of the
minimum and maximum predicted values, so if the lower bound is at 1,
then upper bound should be considered as an upper limit. In these plots,
any lower limits are indicated by arrows. The binary phase is used for the x-axis.
There are 10 cases
showing a finite range for the O $R_{fir}$. There are two O and Ne
$R_{fir}$ values at phase $\approx 0.65$, but in different binary
orbits, indicating the same finite radial locations (within the errors)
for each ion (O at $\approx 7-9$, and Ne at $\approx 4-6$), which could
mean that at least for O and Ne the behavior is repeatable. This is not
seen in Mg or Si. For Mg, in one case for $\phi\approx$0.65 there
is a finite $R_{fir}$ ($\approx$2), whereas the other case at
$\phi\approx$0.65 indicates only a lower bound of $\approx$4.5.  For Si
there are five $R_{fir}$ with finite ranges, all within the errors of
one another. Si has four $R_{fir}$ at
$\approx 1$ since the observed $f/i$ for these phases were below their
respective minimum $f/i$.  This behavior suggests that these regions
producing the majority of the higher energy emission lines may be
experiencing significant dynamic fluctuations in density and/or
temperature.

\begin{figure}[!htb] \centering
\includegraphics[width=\columnwidth]{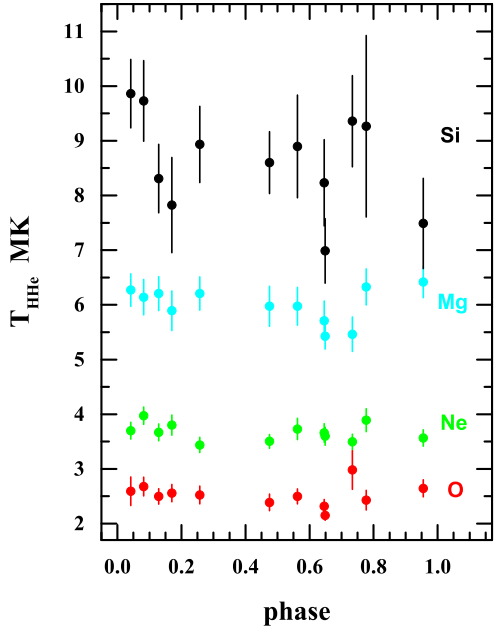}
\caption{Phase dependence of the Si, Mg, Ne, and O T$_{HHe}$ calculated from the H/He
ratios of each of the twelve 40 ks time-sliced spectra.}\label{tx}
\end{figure}

The 12 derived H/He temperatures ($T_{HHe}$) versus phase for each ion
are shown in Fig.\ \ref{tx}. In general, for all ions there
is very little variation in $T_{HHe}$ with phase, although one could
easily argue that at certain phases there are minor fluctuations.

A verification of these results was obtained using the Potsdam
Wolf-Rayet (PoWR) code \citep{hamann04} to perform a similar $f/i$
analysis that included diffuse wind emission and limb darkening.
Differences in the results obtained using the two methods were
negligible compared to measurement uncertainties.

\subsection{Non-detection of Stellar Wind Occultation Effects}

One goal of this program was to use the variable occultation of the
primary wind by the essentially X-ray-dark secondary, \delOri\ Aa2, as
it orbits the primary, mapping  the ionization, temperature, and
velocity regimes within the primary's stellar wind.  The secondary star,
\delOri\ Aa2, is orbiting deep within the wind of the primary star,
\delOri\ Aa1.  Based on the binary separation of 2.6 R$_{Aa1}$ and our
calculations of the $fir$-inferred radii of the various ions, we expect
the secondary to be outside the onset radius of S emission in the
primary wind, very close to or inside of the onset radius of Si, and
inside the onset radii of Mg, Ne, and O.  We can make a simple model of
the expected lightcurves for these emission lines for \delOri\ Aa.   If
the secondary is outside the onset radius of an ion, the lightcurve will have a
maximum and relatively constant flux value between $\phi$ $\approx$0.25
and $\approx$0.75.  The lightcurve will have a relatively constant but
lower flux at $\phi$ $\approx$0.75-0.25 as it occults both the back and
front sides of the onset-radius shell.  If the secondary is inside the
onset radius of an ion, then the lightcurve will be at a maximum flux near $\phi$=
0.0 and 0.5.  Between these two phases, the lightcurve will have a lower
and relatively constant flux value as it occults only the back side of
the onset-radius shell.

The $i$ line  of the He-like triplet is formed deep in the wind while
both the $r$ and $f$ lines are more distributed throughout the wind.
Our best chance of identifying occultation effects is probably from the
fluxes of the $i$ lines, as the  shell of $i$-line emission will be
thinner than either the $f$-line emission shell or the $r$-line emission
shell. To estimate any occultation effect that we might see in these
lightcurves, we calculated the maximum volume of $i$-line emission that
could be occulted by the secondary using various parameters.  The total $i$-line emission expected from an emitting shell around the surface of the star was estimated using the spherical volume of the shell.
The secondary star is assumed to have a radius of 0.3  R$_{star}$.
The percentage of $i$-line emission occulted by the secondary star will be maximum when the largest column of emitting material is occulted, which is at the rim of the shell.  An estimate of the occulted emission was based on the volume of the spherical cap of the emitting shell, with a height equal to the diameter of the secondary star.  Care was taken to subtract the volume of the star that might be included in the cap.
We find
that, for the extreme case of an  $i$-line emission shell at the surface
of the primary star with a thickness of  0.01 R$_{Aa1}$, the maximum
occultation of the $i$-line flux would reduce the flux by $\approx$20\%.  If the
thickness of the $i$-line shell is instead a more likely value of 0.1
R$_{Aa1}$ but still in contact with the surface of the star, the flux of
the $i$ emission line would be reduced by only about 3\% due to
occultation by the secondary.  In these estimates, the secondary would
not be at primary minimum $\phi$=0.0, but instead projected on the rim
of the $i$-line emission shell where the column of $i$-line emission is
greatest. A thicker $i$-line shell would reduce the amount of
occultation further because of the finite size of the projected
secondary star in relation to the volume of the sphere of $i$-line
emission.  Any other position of the secondary in the orbit, or any
larger shell of $i$-line emission, would reduce the percentage of
occultation.

\begin{figure}[!htb]
\centering
\includegraphics[width=0.9\columnwidth]{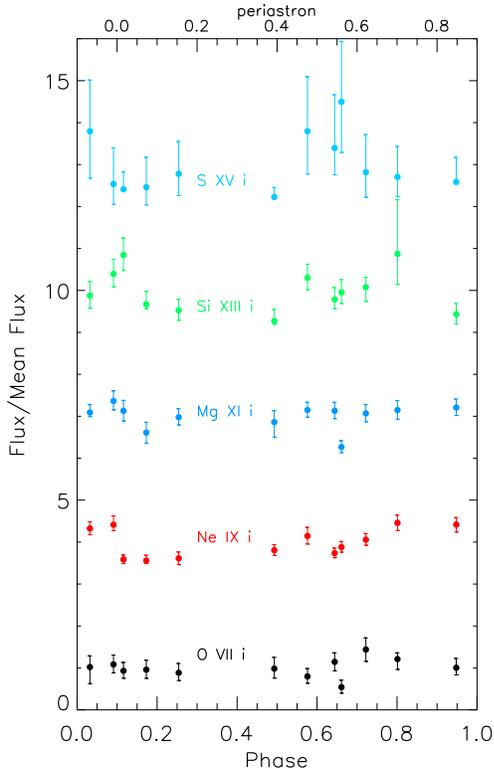}

  \caption{Flux of He-like $i$ component of the triplet based on
  Gaussian fits vs phase.  Errors are 1$\sigma$ confidence limits.
  Fluxes have been normalized by dividing by the mean flux for the
  respective line.  Phase with respect to periastron is indicated at the
  top of the plot.   The plot for each ion is offset by a value of 3
  from the previous one for clarity.} \label{fig:i-s}
  \end{figure}

Fig.\ \ref{fig:i-s} shows the flux measurements for the $i$ lines of the
He-like triplets.  We have previously found a linear increase in flux
over the time period of the observations which is not removed in these
plots.    While the estimates of the fluxes from the \sfir-$i$ and
\sifir-$i$ lines that might have an onset radius very close to the
stellar surface of the primary star do not preclude the existence of
occultation effects, we unfortunately cannot identify such variability
which would be at the 1-2\% level, particularly due to the other
identified variations on the order of 10-15\% and the errors of the
measurements.

\section{DISCUSSION} \label{section:discussion}

\subsection{Effects of a Wind-Star Collision} \label{section:cw}

An important and unexpected result of this analysis is the discovery of
variability of  line widths  with binary phase. H-like emission line
widths are at a minimum at $\phi$=0.0 when the secondary is in front of
the primary, and to a lesser degree at $\phi$=0.7.  The line widths are
at a maximum near $\phi$= 0.2 and $\phi$=0.8, close to quadrature. The
phase-dependent variability of the emission line widths must therefore
be related to interaction between the primary and the secondary.

\begin{figure*}[!htb] \centering \includegraphics[height=60mm]{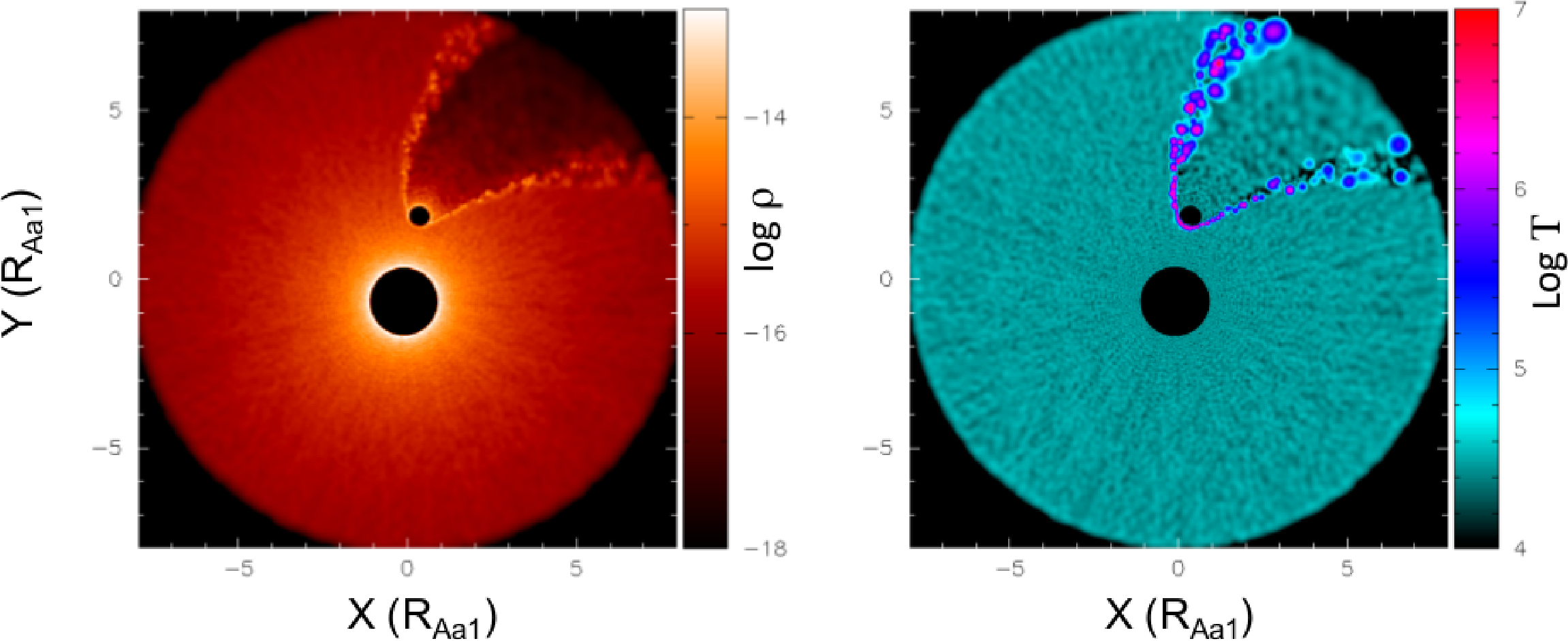}
  \caption{Model of density and temperature structure of the binary
  orbital plane of the SPH simulation of Aa1 (larger black circle) and
  Aa2 (smaller black circle), based the parameters from Paper IV and the
  model in Paper I. The collision of the wind of the primary star
  against the secondary star produces a low density cavity within the
  primary wind.  The perimeter of the low density cavity is a shocked
  bow shock of higher density than either star's wind region (left
  panel).  In the temperature plot on the right, only the hot gas from
  along the wind-collision, bow shock boundary is shown since the SPH
  simulation does not include the embedded wind shocks of either delta
  Ori Aa1 or delta Ori Aa2.  The primary wind collides directly with the
  secondary surface in this simulation primarily because of the large
  difference in mass loss rate ($\dot{M}_{Aa1}/\dot{M}_{Aa2}\approx
  40$). }
  \label{fig:temp} \end{figure*}

In Paper I we developed a model to represent the effect of the secondary
star on the wind region of the primary.   Our 1D, line-of-centers, CAK
calculations presented in Paper I showed that radiative braking does not
occur in this system, so the primary wind directly impacts the surface
of the secondary star.  A similar example has been observed for CPD
$-41^{\circ}$7742, along with an eclipse of the X-ray emitting colliding
wind region when the two stars were perfectly aligned \citep{sana05}.  A
3D smoothed particle hydrodynamics (SPH) code
\citep{Russell13,MaduraP13} was then used to simulate the effect of the
wind-wind collision in \delOri\ Aa.   The colliding winds form an
approximately  cone-shaped cavity  in the wind of the primary with the
secondary star at the apex of the cone.  The cavity has a half-opening
angle of $\approx$30\degr, so the solid angle fraction taken up by the
cone is $\approx$8\%.  A bow shock surrounds the cavity, and within this
cavity the secondary wind prevails, yielding lower densities as well as
very little X-ray flux.  Fig.\ \ref{fig:temp}, reproduced from Paper I,
shows the density and temperature structure of the winds and their
interaction in the binary orbital plane.  The hot gas in the shock at
the interface between the lower density cavity and the primary star wind
was calculated in Paper I to produce at most 10\% of the observed X-ray
flux. According to this model, as the cavity rotates around the primary
star from the blue- to red-shifted part of the wind, emission line
profiles should change in shape and velocity  (as long as the radius of
line formation is similar to or larger than the location of the apex of
the bow shock).

%
%
%
\begin{figure*}[!htb] \centering
\includegraphics{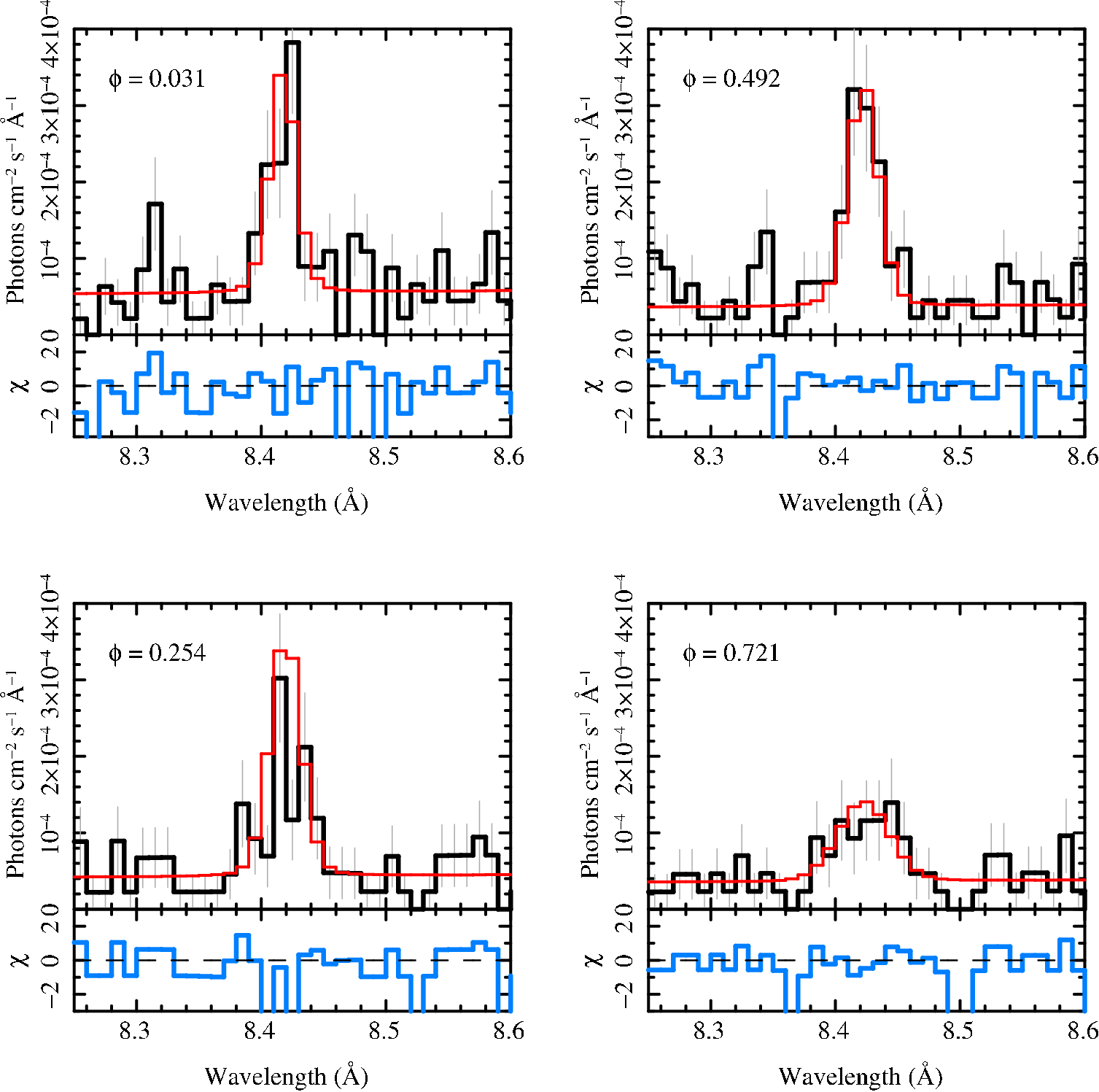}
\caption {\mgH\ profile  overplotted with
the Gaussian fit.  Upper panel left: Phase is centered at 0.031; Upper
panel right: Phase is centered at 0.492; Lower panel left: Phase is
centered at 0.254; Lower panel right: Phase is centered at
0.721.}\label{fig:mg12_conj}
\end{figure*}

The variability of the emission-line widths we have observed may
potentially be explained by this cavity in the primary wind caused by
the wind interaction with the secondary.  When viewed at $\phi$=0.0, the
cavity will occupy a region of the primary stellar wind that would
otherwise be the formation region of  emission with high negative
velocities.  The emission line profiles viewed at this phase might then
be truncated at the largest negative velocities, creating a
comparatively narrower profile than one expects without the presence of
the secondary.  At $\phi$=0.5, some of the most positive velocities
would not be detected in the emission-line profiles.  Wind absorption
must be taken into account at $\phi$=0.5 and the effect of the cavity
may be less pronounced.   Such emission truncation at $\phi$=0.0 and 0.5
is suggested in Fig.\ \ref{fig:mg12_conj}, displaying the  \mgH\
profiles and fits for  the time-sliced spectrum near $\phi$=0.0 and the
time-sliced spectrum near $\phi$=0.5.  The \mgH\ line was chosen as a
relatively strong line that was also used in the template analysis in
Sect.\ \ref{section:composite}.   The red line in the figure is the
Gaussian fit to the line, the black line is the time-sliced spectrum for
a particular phase interval, and the lower panel of each plot shows the
$\delta\chi$ statistic (though the Cash statistic was actually used in
the fitting). Negative velocities appear somewhat under-represented in
the time-sliced spectrum near $\phi$=0.0, although the profile near
$\phi$=0.5 does not appear to be asymmetric. There could be other
explanations  for this characteristic, such as velocity changes in the
centroid or non-Gaussian profiles.  At quadrature, $\phi$=0.25 and 0.75,
the velocities normally produced in the embedded wind of the primary and
replaced by the region occupied by  the cavity will tend to be near zero
velocity,  resulting in emission lines of expected widths, but somewhat
non-Gaussian peaks, such as flat-topped or skewed peaks.  This
prediction is not inconsistent with the profiles observed near
quadrature in the \delOri\ Aa time-sliced spectra, which are broader
than the profiles seen near conjunction (see Fig.\ \ref{fig:mg12_conj}
as an example).

\begin{figure}[!htb]
\centering
\includegraphics[scale=0.4]{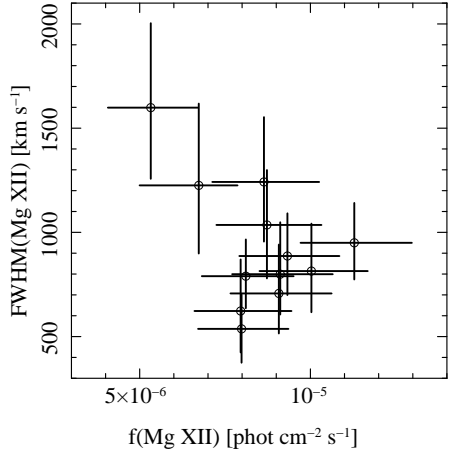}
\caption{Gaussian parameters for \mgH\
lines for each of the 12 time-sliced spectra.  Error bars are 1$\sigma$.
Flux vs FWHM.}\label{fig:mg12_dave}
\end{figure}

However, further analysis of the \mgH\ line reveals a more complex
scenario. Fig.\ \ref{fig:mg12_dave} shows the correlation of FWHM with
flux for \mgH\ for the 12 time-sliced spectra, based on the Gaussian
fits.  While most of the points show a similar distribution with little
or no dependence of FWHM on flux, the three points from the first
\chandra\ observation, \obs 14567, have significantly larger FWHM
values along with lower flux.  We have carefully checked that there is no known reason to
expect this result to be instrumental. We conclude that the variability
in the \mgH\ line has several time scales, of which the binary orbit and
colliding winds associated with the binary system are only one
component.

%
%
%
\subsection{Stellar pulsations and Corotation Interacting Regions}

Pulsations have been found in only a few O stars, but those that have
been found are mostly in late-O stars, such as $\zeta$ Oph, O9.5V
\citep{walker05}.  Variations due to pulsations are cyclic, even over
long timescales
\citep[i.e.][]{prinja86,kaper96,kaper97,kaper99,massa95,prinja88,
henrichs88}. Models predict that massive stars near the main sequence
will experience pulsations of several types, including non-radial
pulsations (NRP), $\beta$ Cep instabilities, and {\it l}-mode pulsations
\citep{cox92,pamy99}.

Another source of variability relates to phenomena near or on the
surface of the star, such as bright spots, which tend to be transient
over a few stellar rotations or less.  For many  years DAC variability
in UV P-Cygni profiles has been recognized, and is believed to be common
in O stars.  \citet{cranmer96}  provided a model of ad hoc photospheric
perturbations in the form of bright spots that was able to reproduce the
DAC phenomenon.  CIRs, possibly related to DACs, are perturbations that
start at the base of the stellar wind and can extend far out into the
wind, and thus are tied to the rotation period of the star, although
they could also be transient.  Presumably there can be multiple CIRs
distributed over  the surface of the star, cumulatively resulting in
variations  that appear to be shorter than the rotation period, although
each individual spot and its associated CIR will rotate with the star.

The periods we have identified in the \chandra\ lightcurve of \delOri\
Aa (4.76d and 2.04d,  see Sect.\ \ref{section:lc}), are possibly the
X-ray signature of  CIRs or pulsations.  The 4.76$\pm$0.3d
period may be the rotation period of the primary star.  Based on a $v
~sin~ i$ value of 130 \ks\ (Paper IV) and the estimated value of
R$_*$=12\Rsun , and also including the assumption of alignment between
the rotational and orbital axes,  the rotational period should be about
4.7d, consistent with our strongest period of 4.76d.   A single
non-transient CIR  would share the stellar rotational period.
 We did not find evidence of  the binary orbital period of 5.73d in the
X-ray lightcurve, indicating that most of the X-ray variability is not
related to the orbital motion, at least for the limited orbital data we
have.

There is an apparent increase in flux over 9 days, suggesting long-term
variability.  We do not have a sufficient time baseline to quantify this
component of variability, but it is clearly not associated with the
binary period.   We suggest this long-term variability is due to
pulsations such  as NRP and/or increasing CIR activity.

\section{CONCLUSIONS}\label{section:conclusions}

\chandra\ high-resolution grating X-ray spectra of  \delOri\ Aa acquired
in 2012 with a total exposure time of $\approx$479 ks have been analyzed
for phase-resolved and time-resolved variability. Several components of
variability were detected in our analyses.

 The count rate of the entire spectral range increased during the 9-day
 observing campaign by approximately 25\%.   We cannot constrain the
 cause of this longer-term variability with this dataset, but we
 speculate this may be related to stellar pulsations or CIRs and other
 wind instabilities.

An important result of the period searches in the X-ray data is the
binary motion seems to contribute very little to the variability of the
total flux. A period search of the total X-ray flux lightcurve yielded
periods of 4.76$\pm$0.3d  and 2.04$\pm$0.5d after removal of the
long-term trend, both of which are less than the binary period of 5.73d.
A period search including the early 2001 \chandra\ observation as well
as the 2012 data gave a period near 5.0d, within the errors of the 4.76d
period determined from the normalized 2012 spectra. The 4.76d period is
consistent with the secondary period found by MOST of 4.614d; thus it is
present in both X-ray and optical data.  We suggest that this may be
the rotation period of \delOri\ Aa1 based on estimates of $v~sin~i$ and
the radius of \delOri\ Aa1.  The 2.04d period, also found  in the MOST
photometry, may be associated with pulsations or CIRs.

 Flux variability of individual emission lines was confirmed with
 statistical tests for the He-like triplets of  \sfir, \sifir, and
 \nefir\ (contaminated with an \ironsevt\ line), as well as the
 \irontwenty\ complex.  Also, several line profiles are apparently
 non-Gaussian with blue-shifted centroids of about -80 \ks\ prevalent,
 possibly indicating that line-fitting with wind profiles would be more
 appropriate. Derived $R_{fir}$ are in similar
 ranges for O stars of the spectral type of \delOri\ Aa1.

 For first time, phase-dependent variability in the X-ray emission line
 widths has been found  in a binary system.  Line widths are at a
 minimum at $\phi$=0.0 and at a maximum at $\phi$=0.2 and 0.8,
 approximately.  It is thus likely that the line widths are dependent on
 an interaction between the primary and secondary.  The variation could
 qualitatively be explained as the result of a cavity in the primary
 wind produced by wind-wind collision.  According to this model, the
 cavity created by the colliding winds would be of comparatively lower
 density, causing a reduction in blueward or redward emission at
 conjunctions, and in principle making the lines narrower at
 conjunctions than at quadratures of the binary phase.  The spectra
 presented in this paper are possibly consistent with this idea,
 although additional short-term variability of the line widths is
 suggested.

One goal of the 2012 \chandra\ observing program of \delOri\ Aa was  to
allow observations of a massive star stellar wind as the short-period
secondary occulted different regions of emission formation on its
journey around the primary star.  We predict the reduction in the flux
levels due to occultation to be about 1-3\% at most.  Additional
variability from other sources of greater magnitude, as well as limited
signal-to-noise in the data, make it impossible to identify occultation
in our dataset at such a low percentage when there are clearly
variations in the 10-15\% range. In particular, detailed analysis of
\mgH\ showed that flux, radial velocity, and FWHM vary  both within a
single orbit and within the dataset as a whole.

The variability we see in the emission from \delOri\ Aa is probably a
composite of several effects, including the long-term, greater than
9-day, photometric variability, binary orbit FWHM effects, inter-orbit
variability and intra-orbit variability. It is likely that CIRs and/or
pulsations play an important role in the variability. New long
observations with the higher sensitivities offered by XMM-Newton would
probably help resolve some of the photometric issues. Questions remain
concerning the source of the periods, phase-dependency of line profiles,
various time scales of variability, and detailed modeling of the line
width variability. Chandra observations at specific phases, such as
conjunction and quadrature, and with a longer timeline, would be useful
in verifying the model as well as parameterizing the variability we have
seen. Additional analysis of the UV DACs may clarify the sources of some
of the components of the variability and in particular the rotation
period of \delOri\ Aa.

\section{Acknowledgements}

The authors acknowledge the constructive comments of the anomymous referee.  MFC, JSN, WLW, CMPR, and KH are grateful for support provided by the
National Aeronautics and Space Administration through Chandra Award
Number GO3-14015A, G03-14015E, and GO3-14015G issued by the Chandra
X-ray Observatory Center, which is operated by the Smithsonian
Astrophysical Observatory for and on behalf of the National Aeronautics
Space Administration under contract NAS8-03060. DPH was supported by
NASA through the Smithsonian Astrophysical Observatory contract
SV3-73016 to MIT for the Chandra X-Ray Center and Science Instruments.
YN acknowledges support from the Fonds National de la Recherche
Scientifique (Belgium), the Communaut\'e Fran\c caise de Belgique, the
PRODEX XMM and Integral contracts, and the ‘Action de Recherche
Concert\'ee (CFWB-Acad\'emie Wallonie Europe). AFJM is grateful for
financial aid from NSRC (Canada) and FRQNT (Quebec).  NDR gratefully
acknowledges his CRAQ (Centre de Recherche en Astrophysique du Qu\'ebec)
fellowship.  LMO acknowledges support from DLR grant 50 OR 1302.  NRE is
grateful for support from the Chandra X-ray Center NASA Contract
NAS8-03060.  JLH acknowledges support from NASA award NNX13AF40G and NSF
award AST-0807477.  MFC, JSN, and KH also acknowledge  helpful
discussions with John Houck and Michael Nowak on data analysis with
ISIS, and Craig Anderson for technical support.   This research has made
use of data and/or software provided by the High Energy Astrophysics
Science Archive Research Center (HEASARC), which is a service of the
Astrophysics Science Division at NASA/GSFC and the High Energy
Astrophysics Division of the Smithsonian Astrophysical Observatory. This
research made use of the Chandra Transmission Grating Catalog and
archive (http://tgcat.mit.edu).  This research also has made use of
NASA's Astrophysics Data System.

Facilities: \facility{CXO}, {MOST}
\bibliographystyle{apj} \bibliography{deltaori}

\input TABLE_HFIR_SI_6A.tex
\input TABLE_HFIR_MG_6A.tex
\input TABLE_HFIR_NE_6A.tex
\input TABLE_HFIR_O_6A.tex

\end{document}

%% file: parameters.tex


\begin{deluxetable}{ll}
\tabletypesize{\scriptsize}
\tablecaption{System  parameters for \del\ Ori Aa1+Aa2.\label{table:params}}
\tablewidth{0pt}
\tablecolumns{2}
\tablehead{
\colhead{Parameter} &
\colhead{Value}

}
\startdata



\hline
Sp. Type (Aa1) & O9.5II \tablenotemark{a,b,d}  \\
Sp. Type (Aa2) & B1V \tablenotemark{a}\\
 \hline
 $ D $ [pc]  &  $380$ (adopted)\tablenotemark{a}  \\
\hline
$ R [R_\odot] $ (Aa1) &  $16.5\pm 1$    \\
$ R [R_\odot] $ (Aa2) &  $6.5^{+2}_{-1.5}$      \\
\hline
\hline
  {\bf Binary Period \tablenotemark{b}} & \\
\hline
 $ P [d] $  & $ 5.732436^{d}$    \\
 $E_{0}$ (primary min, HJD)   & $ 2456277.790\pm0.024$     \\
 $T_{0}$ (periastron, HJD) & $ 2456295.674\pm0.062$ \\
 $ a [\rsun] $  &  $ 43.1 \pm 1.7 $   \\
 $ i $ [deg.]  & $76.5  \pm 0.2 $    \\
 $\omega$ [deg.] & $141.3 \pm 0.2$  \\
 $\dot{\omega}$ [deg. yr$^{-1}$]  & $ 1.45 \pm 0.04 $ \\
 $ e $  & $0.1133 \pm 0.0003$    \\
 $\gamma$~[\ks] &   $15.5 \pm 0.7 $ \\
Periastron-based $\phi$ & 0.116+Photometric-based $\phi$ \\
 \hline
\hline
 {\bf MOST optical secondary periods [d]}  & \\
\hline
$MOST_{F1}$ &  $2.49 \pm 0.332$  \\

$MOST_{F2}$  & $4.614 \pm 1.284$  \\

$MOST_{F3}$ &  $1.085 \pm 0.059$  \\

$MOST_{F4}$ &  $6.446 \pm 2.817$  \\

$MOST_{F5}$ &  $3.023 \pm 0.503$  \\

$MOST_{F6}$\tablenotemark{e} &  $29.221 \pm 106.396$  \\

$MOST_{F7}$  & $3.535 \pm 0.707$  \\

$MOST_{F8}$  & $1.01 \pm 0.051$  \\

$MOST_{F9}$ &  $1.775 \pm 0.162$  \\

$MOST_{F10}$ &  $2.138 \pm 0.24$  \\

$MOST_{F11}$ &  $1.611 \pm 0.133$  \\

$MOST_{F12}$ &  $0.809 \pm 0.032$  \\

$MOST_{F13}$ &  $0.748 \pm 0.027$  \\

\enddata
\tablenotetext{a}{ Shenar et al. (2015)}
\tablenotetext{b}{from the low-mass model solution of  Pablo et al. (2015)}
\tablenotetext{c}{\citet{maiz13}}
\tablenotetext{d}{Mayer et al. (2010)}
\tablenotetext{e}{This peak is likely an artifact due to a trend in the data. It is not considered real, but it is formally significant and included in the fit.}
\end{deluxetable}

%% file: obs_table.tex
\begin{deluxetable}{lllllllllr}

\tabletypesize{\footnotesize}
\tablecolumns{10}
\tablewidth{0pc}
\tablecaption{2013 \chandra\ Observations of \del\ Ori Aa\label{tab:obs}}
\tablehead{
\colhead{ObsID}    &  \colhead{Start} &   \colhead{Start} 
& \colhead{End}    &  \colhead{End} &   \colhead{Midpoint}
& \colhead{Midpoint}    &     \colhead{$\Delta T$}
& \colhead{Exposure}    &  \colhead{Roll}   \\
\colhead{} & \colhead{HJD} & \colhead{Phase} & \colhead{HJD} &
\colhead{Phase} & \colhead{HJD} & \colhead{Phase} & \colhead{days} & \colhead{s} &
\colhead{deg.} \\
}

\startdata

14567 & 2456281.21 & 396.604 & 2456282.58 & 396.843 & 2456281.90 & 396.724 & 1.37 & 114982 & 345.2 \\
14569 & 2456283.76 & 397.049 & 2456285.18 & 397.297 & 2456284.47 & 397.173 & 1.42 & 119274 & 343.2 \\
14570 & 2456286.06 & 397.450 & 2456287.52 & 397.705 & 2456286.79 & 397.578 & 1.46 & 122483 &  83.0 \\
14568 & 2456288.67 & 397.905 & 2456290.12 & 398.159 & 2456289.39 & 398.032 & 1.45 & 121988 & 332.7 \\

\enddata


\end{deluxetable}

%% file: period.tex
\begin{deluxetable}{ccc}

\tabletypesize{\scriptsize}
\tablecolumns{3}
\tablewidth{0pc}
\tablecaption{Fourier Periods\label{table:period}}
\tablehead{
\colhead{ID\tablenotemark{a}}    &  \colhead{Period} &   \colhead{Amp.} \\
&d&10$^{-3}$\,ct\,s$^{-1}$ \\
}

\startdata

raw $\cts$ & 5.0$\pm$0.3 & 7.1$\pm$0.7   \\
residual  & 4.76$\pm$0.3 & 4.7$\pm$0.08  \\
prew. res. & 2.04$\pm$0.05 & 3.5$\pm$0.6   \\

\enddata
\tablenotetext {a}{ raw indicates X-ray lightcurve in $\cts$; residual indicates raw lightcurve with linear trend removed; prew. res. indicates raw lightcurve with 4.76d period and linear trend removed. }

\end{deluxetable}

%% file: slice_table.tex
\begin{deluxetable}{rrrrrrr}
\tabletypesize{\scriptsize}
\tablewidth{0pt}
\tablecolumns{7}
\tablecaption{Chandra Time-Sliced Spectra Log\label{table:slice}}
\tablehead{
\colhead{Obsid/slice} &
\colhead{Start HJD} &
\colhead{Start phase} &
\colhead{End HJD} &
\colhead{End phase} &
\colhead{Duration (s)} &
\colhead{Mid phase}

}
\startdata

14567/1 & 56280.718	& 396.606	& 56281.156	& 396.682	& 37811	& 396.644 \\
14567/2	& 56281.156	& 396.682	& 56281.607	& 396.761	& 39000	& 396.722 \\
14567/3	& 56281.607	& 396.761	& 56282.067	& 396.841	& 39693	& 396.801 \\

14569/1 &	56283.267 &	397.051 &	56283.729 &	397.131 &	39948 &	397.091 \\
14569/2 &	56283.729 &	397.131 &	56284.204 &	397.214 &	41000 &	397.173 \\
14569/3 &	56284.204 &	397.214 &	56284.666 &	397.295 &	39906	& 397.254 \\
14570/1 & 	56285.568 &	397.452 &	56286.038 &	397.534 &	40584	 & 397.493 \\
14570/2	 & 56286.038	& 397.534 &	56286.524 &	397.619	& 42000	& 397.576 \\
14570/3 &	56286.524 &	397.619 &	56287.004 &	397.703 &	41521 &	397.661 \\
14568/1	& 56288.177	 & 397.907 & 	56288.648	& 397.989	& 40662	& 397.948 \\
14568/2 &	56288.648	 & 397.989	& 56289.134 &	398.074	& 42000	& 398.032 \\
14568/3	& 56289.134 &	398.074 &	56289.608 &	398.157 &	40941 & 	398.116 \\
\enddata
\end{deluxetable}

%% file: tbl_fprops.tex
\begin{deluxetable}{rrrr}
\tabletypesize{\footnotesize}
\tablewidth{0pt}
\tablecolumns{4}
\tablecaption{TGCat Wavelength Bins\label{table:fprops}}
\tablehead{
\colhead{Label} &
\colhead{$\lambda$} &
\colhead{$\lambda$ low} &
\colhead{$\lambda$ high} \\
\colhead{} &
\colhead{\ang} &
\colhead{\ang} &
\colhead{\ang} 
}
\startdata

c2500\tablenotemark{a}  &  2.50 & 2.00 &  3.00  \\
  S~XVI & 4.75 & 4.70 & 4.80 \\
  c4900 & 4.90 & 4.80 & 5.00 \\
  S~XV & 5.08 & 5.00 & 5.15 \\
  c5700 & 5.70 & 5.40 & 6.00 \\
  Si~XIV & 6.17 & 6.10 & 6.25 \\
  c6425 & 6.42 & 6.30 & 6.55 \\
  Si~III & 6.70 & 6.60 & 6.80 \\
  c7800 & 7.80 & 7.40 & 8.20 \\
  Mg~XII & 8.40 & 8.35 & 8.45 \\
  c8800 & 8.80 & 8.50 & 9.10 \\
  Mg~XI & 9.25 & 9.10 & 9.40 \\
  Fe~XX & 11.20 & 10.40 & 12.00 \\
  Ne~X  &  12.10 & 12.10&  12.20  \\
  c13200  &  13.20 & 13.00 &  13.40  \\
  Ne~IX & 13.60 & 13.40 & 13.80 \\
  Fe~XVII & 15.00 & 14.95 & 15.05 \\
  c14925 & 14.92 & 14.90 & 15.05 \\
 O~VIII & 16.00 & 15.95 & 16.05 \\
  c16450 & 16.45 & 16.20 & 16.70 \\
  Fe~XVII & 17.07 & 17.00 & 17.15 \\
  O~XIII & 18.98 & 18.90 & 19.05 \\
  c20200 & 20.20 & 19.20 & 21.20 \\
  O~VII & 21.85 & 21.50 & 22.20 
\enddata
\tablenotetext {a}{Continuum band labels are \\ a ``c'' followed by the band \\ wavelength in m\AA.  }

\end{deluxetable}

%% file: flux_table_land.tex
\begin{deluxetable}{ccccccccc}
\tabletypesize{\scriptsize}
\tablecolumns{9}
\tablewidth{0pt}
\tablecaption{Emission Line Fluxes\tablenotemark{a}:S and Si.}\label{table:flux:s}
\tablehead{
\colhead{Binary Phase}&\colhead{S XVI}    &  \multicolumn{3}{c}{S XV} & \colhead{Si XIV} & \multicolumn{3}{c}{Si XIII} \\
\cline{3-5} \cline{7-9} \\
\colhead{}&\colhead{} & \colhead{r}   & \colhead{i}    & \colhead{f}  &\colhead{} & \colhead{r}   & \colhead{i}    & \colhead{f}  \\
}

\startdata

0.606-0.682& 1.4$_{-0.8}^{+1.2}$ & 2.6$_{-1.1}^{+1.5}$ & 1.9$_{-0.9}^{+1.7}$ & 1.4$_{-0.8}^{+1.3}$ & 2.6$_{-0.7}^{+0.9}$ & 11.6$_{-1.8}^{+1.8}$ & 3.0$_{-0.8}^{+1.1}$ & 0.8$_{-0.6}^{+0.7}$ \\

0.682-0.761& 0.7$_{-0.1}^{+0.5}$ & 2.6$_{-1.1}^{+1.4}$ & 1.1$_{-0.8}^{+1.2}$ & 1.2$_{-0.8}^{+1.2}$ & 3.7$_{-0.9}^{+1.1}$ & 11.3$_{-1.7}^{+0.8}$ & 4.1$_{-1.3}^{+0.9}$ & 4.5$_{-1.0}^{+1.2}$  \\

0.761-0.841& 0.7$_{-0.1}^{+0.5}$ & 0.3$_{-0.3}^{}$ & 0.9$_{-0.6}^{+1.0}$ & 0.5$_{-0.4}^{+0.7}$ & 3.0$_{-0.9}^{+2.2}$ & 10.4$_{-4.6}^{+3.1}$ & 7.2$_{-2.8}^{+4.9}$ & 5.0$_{-1.1}^{+1.2}$ \\

0.051-0.131&0.1$_{-0.1}^{+0.7}$ & 2.4$_{-1.3}^{+1.2}$ & 0.7$_{-0.6}^{+1.2}$ & 0.7$_{-0.6}^{+1.0}$ & 5.1$_{-1.0}^{+1.2}$ & 12.5$_{-1.7}^{+1.8}$ & 5.3$_{-1.2}^{+1.3}$ & 5.4$_{-1.1}^{+1.2}$ \\

0.131-0.214 &0.6$_{-0.6}^{+0.9}$& 4.2$_{-1.4}^{+1.7}$ & 0.6$_{-0.6}^{+1.0}$ & 5.2$_{-1.5}^{+1.9}$ & 2.8$_{-1.2}^{+1.0}$ & 14.2$_{-1.8}^{+1.8}$ & 2.6$_{-0.4}^{+1.2}$ & 8.7$_{-2.4}^{+2.0}$ \\

0.214-0.295& 0.9$_{-0.7}^{+1.0}$ & 3.1$_{-1.2}^{+1.5}$ & 1.0$_{-0.7}^{+1.0}$  & 1.6$_{-0.9}^{+1.2}$ & 4.3$_{-0.9}^{+1.1}$ & 14.6$_{-1.8}^{+1.8}$ & 2.0$_{-0.9}^{+1.0}$ & 2.9$_{-0.5}^{+0.6}$ \\

0.452-0.534& 2.5$_{-1.1}^{+1.4}$ & 4.2$_{-1.4}^{+0.8}$ & 0.3$_{-0.3}^{}$  & 3.0$_{-1.1}^{+0.7}$ & 4.6$_{-1.0}^{+1.1}$ & 18.6$_{-2.1}^{+1.0}$ & 1.0$_{-1.0}^{}$ & 7.2$_{-1.2}^{+1.3}$ \\

0.534-0.619& 0.9$_{-0.6}^{+1.0}$ & 4.1$_{-1.5}^{+1.8}$ & 2.4$_{-1.4}^{+1.7}$ & 2.1$_{-1.1}^{+1.4}$ & 2.7$_{-0.8}^{+1.0}$ & 9.7$_{-1.4}^{+1.4}$ & 5.0$_{-1.1}^{+1.2}$ & 6.3$_{-1.1}^{+1.2}$ \\

0.619-0.703& 0.6$_{-0.6}^{+0.9}$  & 4.8$_{-1.6}^{+2.1}$ & 3.3$_{-1.6}^{+1.9}$ & 1.6$_{-1.1}^{+1.3}$ & 1.3$_{-0.5}^{+0.6}$ & 12.5$_{-1.6}^{+1.7}$ & 3.7$_{-1.0}^{+1.2}$ & 7.4$_{-1.7}^{+1.5}$  \\

0.907-0.989& 0.7$_{-0.6}^{+1.0}$ & 2.7$_{-0.5}^{+1.4}$ & 0.8$_{-0.8}^{}$ & 2.4$_{-1.1}^{+1.4}$ & 2.2$_{-0.9}^{+1.2}$ & 14.6$_{-1.7}^{+1.9}$ & 1.6$_{-0.9}^{+1.0}$ & 5.8$_{-1.5}^{+1.4}$ \\

0.989-0.074& 0.9$_{-0.7}^{+1.0}$  & 3.1$_{-1.2}^{+1.8}$ & 2.4$_{-1.5}^{+1.6}$ & 1.2$_{-0.8}^{+1.2}$ & 6.8$_{-1.2}^{+1.2}$ & 16.0$_{-1.9}^{+2.0}$ & 3.4$_{-1.2}^{+1.3}$ & 9.1$_{-1.5}^{+1.6}$ \\

0.074-0.157&1.8$_{-0.8}^{+1.2}$ & 6.8$_{-1.7}^{+2.0}$ & 0.6$_{-0.6}^{}$ & 5.0$_{-1.5}^{+1.8}$ & 4.4$_{-1.1}^{+1.3}$ & 19.0$_{-1.4}^{+2.2}$ & 7.0$_{-1.4}^{+1.6}$ & 1.4$_{-1.4}^{}$ \\
\enddata
\tablenotetext{a}{$10^{-6} photons/s/cm^2$.  Listed in time order}
\end{deluxetable}

\begin{deluxetable}{ccccccccc}

\tabletypesize{\scriptsize}
\tablecolumns{9}
\tablewidth{0pt}
\tablecaption{Emission Line Fluxes\tablenotemark{a}: Mg and Ne.}\label{table:flux:mg}

\tablehead{
\colhead{Binary Phase}&\colhead{Mg XII}    &  \multicolumn{3}{c}{Mg XI} & \colhead{Ne X} & \multicolumn{3}{c}{Ne IX} \\
\cline{3-5} \cline{7-9} \\
\colhead{}&\colhead{} & \colhead{r}   & \colhead{i}    & \colhead{f}  &\colhead{} & \colhead{r}   & \colhead{i}    & \colhead{f}  \\
}
\startdata

0.606-0.682 & 7.4$_{-1.4}^{+1.5}$ & 25.4$_{-2.9}^{+3.1}$ & 15.5$_{-2.6}^{+2.8}$ & 9.1$_{-1.9}^{+0.9}$ & 78.3$_{-11.5}^{+9.5}$ & 156.2$_{-17.6}^{+19.0}$ & 77.0$_{-11.6}^{+12.8}$ & 16.1$_{-5.0}^{+8.9}$ \\

0.682-0.761 & 6.4$_{-1.2}^{+1.4}$ & 26.1$_{-3.2}^{+3.4}$ & 14.6$_{-2.7}^{+2.9}$ & 9.7$_{-2.1}^{+2.3}$ & 66.8$_{-9.4}^{+9.8}$ & 167.3$_{-18.2}^{+19.1}$ & 110.4$_{-13.6}^{+15.4}$ & 22.8$_{-7.3}^{+8.5}$ \\

0.761-0.841 & 10.3$_{-1.6}^{+1.5}$ & 23.1$_{-3.2}^{+3.5}$ & 15.7$_{-3.0}^{+3.1}$  & 8.3$_{-2.1}^{+1.7}$ & 69.6$_{-9.5}^{+10.0}$ & 115.1$_{-15.8}^{+16.9}$  & 152.3$_{-19.0}^{+19.3}$ & 28.6$_{-7.4}^{+8.5}$ \\

0.051-0.131  & 10.5$_{-1.6}^{+1.5}$ & 27.0$_{-3.3}^{+3.3}$ & 18.7$_{-2.9}^{+3.3}$ & 7.2$_{-1.6}^{+1.8}$ & 102.8$_{-10.5}^{+10.9}$ & 145.6$_{-13.2}^{+17.9}$ & 147.7$_{-14.7}^{+22.0}$ & 26.4$_{-8.0}^{+7.5}$  \\

0.131-0.214 & 10.0$_{-1.5}^{+1.6}$ & 30.4$_{-3.9}^{+4.4}$ & 8.4$_{-3.4}^{+3.4}$ & 10.5$_{-2.5}^{+2.8}$ & 85.3$_{-13.1}^{+7.2}$ & 136.3$_{-15.0}^{+15.9}$ & 58.3$_{-7.4}^{+13.4}$  & 17.6$_{-7.2}^{+5.5}$ \\

0.214-0.295 & 11.5$_{-1.6}^{+1.7}$ & 28.5$_{-3.3}^{+3.5}$ & 13.4$_{-2.5}^{+2.7}$ & 11.2$_{-2.2}^{+2.5}$ & 81.4$_{-9.6}^{+10.3}$ & 196.0$_{-24.1}^{+25.5}$ & 64.1$_{-15.8}^{+15.6}$ & 15.4$_{-6.1}^{+7.2}$ \\

0.452-0.534 & 9.6$_{-1.5}^{+1.5}$ & 27.8$_{-3.6}^{+4.0}$ & 11.8$_{-4.9}^{+3.8}$ & 10.4$_{-2.4}^{+5.4}$ & 97.9$_{-11.2}^{+10.1}$ & 213.9$_{-21.2}^{+22.7}$ & 84.3$_{-13.1}^{+13.9}$ & 28.0$_{-7.4}^{+8.5}$  \\

0.534-0.619 & 8.8$_{-1.3}^{+1.4}$ & 25.7$_{-2.9}^{+3.1}$ & 15.7$_{-2.4}^{+2.6}$ & 6.6$_{-1.6}^{+1.8}$ & 77.6$_{-9.5}^{+9.8}$ & 147.0$_{-19.8}^{+20.6}$  & 119.7$_{-20.1}^{+21.5}$ & 16.0$_{-7.6}^{+10.6}$ \\

0.619-0.703 & 9.0$_{-1.3}^{+1.4}$ & 37.6$_{-3.8}^{+4.0}$ & 3.6$_{-1.8}^{+2.1}$ & 11.6$_{-2.3}^{+2.5}$  & 70.8$_{-9.1}^{+9.6}$ & 147.7$_{-17.5}^{+19.0}$ & 92.2$_{-12.8}^{+13.9}$ & 36.9$_{-7.7}^{+8.7}$ \\

0.907-0.989& 10.8$_{-1.5}^{+1.6}$ & 23.4$_{-2.9}^{+3.1}$ & 16.5$_{-2.5}^{+2.8}$  & 11.4$_{-2.1}^{+2.3}$ & 80.7$_{-9.7}^{+10.8}$ & 177.5$_{-20.2}^{+21.7}$ & 148.3$_{-18.9}^{+19.0}$  & 30.1$_{-7.7}^{+8.7}$ \\

0.989-0.074 & 9.2$_{-1.3}^{+1.4}$ & 22.3$_{-2.9}^{+2.3}$ & 14.9$_{-1.4}^{+2.5}$ & 8.9$_{-1.8}^{+2.1}$ & 92.9$_{-9.8}^{+11.1}$ & 175.5$_{-17.8}^{+19.6}$ & 138.4$_{-15.7}^{+16.8}$ & 43.7$_{-8.8}^{+9.9}$ \\

0.074-0.157 & 12.3$_{-1.6}^{+1.6}$ & 30.2$_{-3.9}^{+4.4}$ & 15.5$_{-3.3}^{+3.4}$ & 9.4$_{-2.0}^{+2.3}$ & 69.2$_{-9.8}^{+10.2}$  & 157.5$_{-15.9}^{+16.8}$ & 61.2$_{-10.4}^{+11.2}$ & 23.2$_{-6.2}^{+7.2}$ \\
\enddata

\tablenotetext{a}{$10^{-6} photons/s/cm^2$.  Listed in time order}
\end{deluxetable}

\begin{deluxetable}{cccccc}

\tabletypesize{\scriptsize}
\tablecolumns{6}
\tablewidth{0pt}
\tablecaption{Emission Line Fluxes\tablenotemark{a}: O and Fe XVII.\label{table:flux:o}}


\tablehead{
\colhead{Binary Phase}&\colhead{O VIII}    &  \multicolumn{3}{c}{O VII} & \colhead{Fe XVII}  \\
\cline{3-5}  \\
\colhead{}&\colhead{} & \colhead{r}   & \colhead{i}    & \colhead{f}  &\colhead{}  \\
}

\startdata
0.606-0.682  & 1018.6$_{-87.4}^{+93.5}$ & 1101.4$_{-170.4}^{+189.4}$ & 841.2$_{-155.8}^{+158.8}$ & 90.9$_{51.9}^{75.2}$  & 299.0$_{-27.1}^{+24.8}$ \\
0.682-0.761 & 992.1$_{-86.9}^{+91.9}$ & 408.4$_{-121.7}^{+195.2}$ & 1058.2$_{-208.6}^{+203.3}$ & 52.8$_{34.7}^{65.4}$ &  304.3$_{-25.1}^{+27.0}$  \\
0.761-0.841 & 1010.2$_{-112.6}^{+113.3}$ & 913.6$_{-202.3}^{+208.5}$ & 889.8$_{-180.8}^{+110.2}$ & 18.5$_{18.5}^{41.5}$   & 270.7$_{-31.0}^{+33.5}$ \\

0.051-0.131  & 1023.7$_{-86.4}^{+92.6}$ & 666.1$_{-131.6}^{+152.1}$ & 796.8$_{-147.2}^{+164.0}$ & 15.3$_{15.3}^{}$  & 307.8$_{-26.4}^{+28.1}$ \\
0.131-0.214 & 1122.8$_{-88.2}^{+94.8}$ & 845.5$_{-158.0}^{+181.7}$ & 704.1$_{-151.0}^{+167.4}$ & 16.0$_{16.0}^{}$   & 243.1$_{-21.5}^{+22.6}$\\
0.214-0.295 & 920.4$_{-83.1}^{+87.5}$ & 728.1$_{-139.9}^{+154.3}$ & 650.6$_{-136.8}^{+162.3}$ & 15.3$_{15.3}^{}$   & 275.8$_{-23.2}^{+24.7}$\\
0.452-0.534 & 1018.6$_{-86.4}^{+91.3}$ & 990.0$_{-188.5}^{+200.4}$ & 723.5$_{-166.9}^{+197.7}$ & 15.4$_{15.4}^{}$ & 310.2$_{-25.5}^{+27.2}$ \\
0.534-0.619 & 1066.0$_{-86.8}^{+91.4}$ & 880.9$_{-140.9}^{+156.9}$ & 586.3$_{-119.5}^{+137.3}$  & 21.9$_{21.9}^{}$  & 294.5$_{-23.3}^{+24.9}$ \\
0.619-0.703 & 860.9$_{-94.8}^{+68.1}$ & 1215.4$_{-173.6}^{+192.0}$ & 398.6$_{-106.2}^{+122.6}$ & 25.8$_{20.4}^{39.7}$  & 254.7$_{-21.2}^{+24.3}$ \\

0.907-0.989 & 1022.4$_{-86.5}^{+91.5}$ & 689.3$_{-122.0}^{+148.8}$  & 738.6$_{-126.6}^{+162.9}$ & 11.2$_{11.2}^{51.7}$ & 305.2$_{-25.6}^{+27.3}$ \\
0.989-0.074 & 1059.4$_{-86.8}^{+91.6}$ & 683.7$_{-171.0}^{+316.9}$ &
750.0$_{-289.6}^{+197.6}$ & 26.5$_{20.6}^{40.0}$& 321.0$_{-24.1}^{+27.8}$ \\
0.074-0.157 & 1026.2$_{-86.6}^{+91.6}$ & 844.5$_{-140.7}^{+157.8}$ & 684.8$_{-130.2}^{+147.5}$ & 6.6$_{6.6}^{43.9}$ & 375.0$_{-27.7}^{+28.4}$ \\

\enddata

\tablenotetext{a}{$10^{-6} photons/s/cm^2$.  Listed in time order.}

\end{deluxetable}



%% file: TABLE_HFIR_SI_6A.tex
                                                                                                                                  
\begin{deluxetable}{cccccccc}                                                                                                     
\tabletypesize{\footnotesize}                                                                                                     
                                                                                                                                  
\tablewidth{0pt}                                                                                                                  
\tablecolumns{8}                                                                                                                  
                                                                                                                                  
\tablecaption{Silicon Line Ratios and Derived Parameters \label{tab:FISI}}                                                        
\tablehead{                                                                                                                       
\colhead{phase}          &                                                                                                        
\colhead{MJD}            &                                                                                                        
\colhead{$f/i$ ratio}    &                                                                                                        
\colhead{$G$ ratio}      &                                                                                                        
\colhead{$H/He$ ratio}   &                                                                                                        
\colhead{$R_{fir}/R{*}$} &                                                                                                        
\colhead{$T_{G}$  MK}    &                                                                                                        
\colhead{$T_{HHe}$  MK}}                                                                                                          
\startdata                                                                                                                        
                                                                                                                                  
\multicolumn{8}{c}{\textit{\underline{Time Ordered}}} \\ \\                                                                       
                                                                                                                                  
   .646 &56280.93 & $ 0.52\pm 0.26$ & $ 0.41\pm 0.12$ & $ 0.25\pm 0.09$ & $ < 1.02$       &   \nodata       & $ 8.23\pm 0.79$ \\ 
   .734 &56281.38 & $ 1.30\pm 0.46$ & $ 0.83\pm 0.17$ & $ 0.38\pm 0.11$ & $ 1.46\pm 0.41$ & $ > 4.50$       & $ 9.36\pm 0.83$ \\ 
   .777 &56281.84 & $ 0.73\pm 0.37$ & $ 1.47\pm 0.72$ & $ 0.41\pm 0.24$ & $ 1.10\pm 0.09$ & $ 6.45\pm 3.99$ & $ 9.27\pm 1.66$ \\ 
   .082 &56283.49 & $ 1.19\pm 0.35$ & $ 0.94\pm 0.19$ & $ 0.44\pm 0.11$ & $ 1.32\pm 0.28$ & $ 7.16\pm 3.36$ & $ 9.73\pm 0.74$ \\ 
   .170 &56283.97 & $ 3.21\pm 1.14$ & $ 0.87\pm 0.20$ & $ 0.20\pm 0.09$ & $ > 2.5$        & $10.20\pm 6.18$ & $ 7.82\pm 0.87$ \\ 
   .257 &56284.44 & $ 1.99\pm 0.96$ & $ 0.42\pm 0.09$ & $ 0.32\pm 0.08$ & $ > 1.1$        &   \nodata       & $ 8.93\pm 0.70$ \\ 
   .475 &56285.79 &   \nodata       & $ 0.49\pm 0.09$ & $ 0.28\pm 0.07$ &   \nodata       &   \nodata       & $ 8.60\pm 0.57$ \\ 
   .562 &56286.28 & $ 1.42\pm 0.40$ & $ 1.25\pm 0.25$ & $ 0.32\pm 0.11$ & $ 1.56\pm 0.42$ & $ 3.73\pm 0.83$ & $ 8.90\pm 0.94$ \\ 
   .649 &56286.77 & $ 2.19\pm 0.77$ & $ 0.94\pm 0.19$ & $ 0.12\pm 0.05$ & $ > 1.5$        & $ 7.13\pm 3.38$ & $ 6.99\pm 0.59$ \\ 
   .955 &56288.40 & $ 3.95\pm 2.29$ & $ 0.57\pm 0.14$ & $ 0.17\pm 0.08$ & $ > 1.7$        & $ > 12$         & $ 7.49\pm 0.83$ \\ 
   .042 &56288.89 & $ 2.91\pm 1.14$ & $ 0.83\pm 0.16$ & $ 0.46\pm 0.10$ & $ > 1.9$        & $11.11\pm 6.39$ & $ 9.86\pm 0.63$ \\ 
   .129 &56289.38 & $ 0.99\pm 0.22$ & $ 0.73\pm 0.11$ & $ 0.25\pm 0.07$ & $ 1.14\pm 0.13$ & $ > 8$          & $ 8.31\pm 0.63$ \\

\enddata                                                                                                                          
\tablecomments{Null entries imply unresolved ratio and/or parameter ranges.}                                                      
\end{deluxetable}                                                                                                                 

%% file: TABLE_HFIR_MG_6A.tex
                                                                                                                                  
\begin{deluxetable}{cccccccc}                                                                                                     
\tabletypesize{\footnotesize}                                                                                                     
                                                                                                                                  
\tablewidth{0pt}                                                                                                                  
\tablecolumns{8}                                                                                                                  
                                                                                                                                  
\tablecaption{Magnesium Line Ratios and Derived Parameters \label{tab:FIMG}}                                                      
\tablehead{                                                                                                                       
\colhead{phase}          &                                                                                                        
\colhead{MJD}            &                                                                                                        
\colhead{$f/i$ ratio}    &                                                                                                        
\colhead{$G$ ratio}      &                                                                                                        
\colhead{$H/He$ ratio}   &                                                                                                        
\colhead{$R_{fir}/R{*}$} &                                                                                                        
\colhead{$T_{G}$  MK}    &                                                                                                        
\colhead{$T_{HHe}$  MK}}                                                                                                          
\startdata                                                                                                                        
                                                                                                                                  
\multicolumn{8}{c}{\textit{\underline{Time Ordered}}} \\ \\                                                                       
                                                                                                                                  
   .646 &56280.93 & $ 0.54\pm 0.13$ & $ 0.94\pm 0.16$ & $ 0.32\pm 0.07$ & $ 2.30\pm 0.34$ & $ 4.35\pm 1.64$ & $ 5.71\pm 0.36$ \\ 
   .734 &56281.38 & $ 0.66\pm 0.19$ & $ 0.93\pm 0.18$ & $ 0.27\pm 0.06$ & $ 2.61\pm 0.49$ & $ 4.82\pm 2.12$ & $ 5.46\pm 0.31$ \\ 
   .777 &56281.84 & $ 0.51\pm 0.15$ & $ 1.02\pm 0.21$ & $ 0.48\pm 0.10$ & $ 2.20\pm 0.41$ & $ 3.86\pm 1.50$ & $ 6.33\pm 0.33$ \\ 
   .082 &56283.49 & $ 0.38\pm 0.11$ & $ 0.96\pm 0.18$ & $ 0.42\pm 0.08$ & $ 1.85\pm 0.31$ & $ 4.20\pm 1.63$ & $ 6.14\pm 0.32$ \\ 
   .170 &56283.97 & $ 1.26\pm 0.60$ & $ 0.62\pm 0.16$ & $ 0.36\pm 0.07$ & $ 4.55\pm 1.94$ & $36.40\pm30.34$ & $ 5.89\pm 0.36$ \\ 
   .257 &56284.44 & $ 0.83\pm 0.23$ & $ 0.86\pm 0.16$ & $ 0.44\pm 0.08$ & $ 3.10\pm 0.62$ & $17.48\pm14.41$ & $ 6.21\pm 0.30$ \\ 
   .475 &56285.79 & $ 1.05\pm 0.53$ & $ 0.82\pm 0.23$ & $ 0.38\pm 0.08$ & $ 3.73\pm 1.49$ & $25.29\pm22.37$ & $ 5.97\pm 0.37$ \\ 
   .562 &56286.28 & $ 0.42\pm 0.13$ & $ 0.87\pm 0.15$ & $ 0.37\pm 0.07$ & $ 1.97\pm 0.36$ & $16.95\pm13.89$ & $ 5.97\pm 0.35$ \\ 
   .649 &56286.77 & $ 3.05\pm 1.67$ & $ 0.41\pm 0.09$ & $ 0.26\pm 0.05$ & $ 4.54\pm 0.00$ & $ > 50$         & $ 5.43\pm 0.24$ \\ 
   .955 &56288.40 & $ 0.68\pm 0.17$ & $ 1.19\pm 0.21$ & $ 0.50\pm 0.10$ & $ 2.71\pm 0.46$ & $ 2.68\pm 0.64$ & $ 6.42\pm 0.28$ \\ 
   .042 &56288.89 & $ 0.57\pm 0.14$ & $ 1.11\pm 0.18$ & $ 0.46\pm 0.09$ & $ 2.38\pm 0.37$ & $ 3.02\pm 0.75$ & $ 6.27\pm 0.30$ \\ 
   .129 &56289.38 & $ 0.60\pm 0.19$ & $ 0.82\pm 0.17$ & $ 0.44\pm 0.08$ & $ 2.45\pm 0.50$ & $20.51\pm17.27$ & $ 6.21\pm 0.31$ \\

\enddata                                                                                                                          
\end{deluxetable}                                                                                                                 

%% file: TABLE_HFIR_NE_6A.tex
                                                                                                                                  
\begin{deluxetable}{cccccccc}                                                                                                     
\tabletypesize{\footnotesize}                                                                                                     
                                                                                                                                  
\tablewidth{0pt}                                                                                                                  
\tablecolumns{8}                                                                                                                  
                                                                                                                                  
\tablecaption{Neon Line Ratios and Derived Parameters \label{tab:FINE}}                                                           
\tablehead{                                                                                                                       
\colhead{phase}          &                                                                                                        
\colhead{MJD}            &                                                                                                        
\colhead{$f/i$ ratio}    &                                                                                                        
\colhead{$G$ ratio}      &                                                                                                        
\colhead{$H/He$ ratio}   &                                                                                                        
\colhead{$R_{fir}/R{*}$} &                                                                                                        
\colhead{$T_{G}$  MK}    &                                                                                                        
\colhead{$T_{HHe}$  MK}}                                                                                                          
\startdata                                                                                                                        
                                                                                                                                  
\multicolumn{8}{c}{\textit{\underline{Time Ordered}}} \\ \\                                                                       
                                                                                                                                  
   .646 &56280.93 & $ 0.23\pm 0.10$ & $ 0.60\pm 0.11$ & $ 0.82\pm 0.12$ & $ 3.93\pm 0.96$ &   \nodata       & $ 3.66\pm 0.17$ \\ 
   .734 &56281.38 & $ 0.21\pm 0.08$ & $ 0.80\pm 0.13$ & $ 0.70\pm 0.10$ & $ 3.66\pm 0.73$ & $ > 2.7$        & $ 3.49\pm 0.14$ \\ 
   .777 &56281.84 & $ 0.19\pm 0.06$ & $ 1.55\pm 0.28$ & $ 1.01\pm 0.17$ & $ 3.54\pm 0.61$ & $ 1.05\pm 0.17$ & $ 3.89\pm 0.21$ \\ 
   .082 &56283.49 & $ 0.17\pm 0.05$ & $ 1.19\pm 0.18$ & $ 1.07\pm 0.14$ & $ 3.35\pm 0.60$ & $ 1.63\pm 0.49$ & $ 3.97\pm 0.16$ \\ 
   .170 &56283.97 & $ 0.27\pm 0.11$ & $ 0.56\pm 0.11$ & $ 0.92\pm 0.13$ & $ 4.24\pm 1.00$ &   \nodata       & $ 3.80\pm 0.18$ \\ 
   .257 &56284.44 & $ 0.25\pm 0.12$ & $ 0.40\pm 0.10$ & $ 0.66\pm 0.10$ & $ 4.02\pm 1.15$ &   \nodata       & $ 3.44\pm 0.14$ \\ 
   .475 &56285.79 & $ 0.33\pm 0.11$ & $ 0.52\pm 0.09$ & $ 0.71\pm 0.09$ & $ 4.86\pm 0.91$ &   \nodata       & $ 3.51\pm 0.13$ \\ 
   .562 &56286.28 & $ 0.14\pm 0.08$ & $ 0.93\pm 0.20$ & $ 0.87\pm 0.14$ & $ 2.89\pm 0.96$ & $ > 1.6$        & $ 3.73\pm 0.19$ \\ 
   .649 &56286.77 & $ 0.40\pm 0.10$ & $ 0.87\pm 0.15$ & $ 0.78\pm 0.12$ & $ 5.39\pm 0.77$ & $ > 2.1$        & $ 3.60\pm 0.17$ \\ 
   .955 &56288.40 & $ 0.21\pm 0.06$ & $ 0.99\pm 0.16$ & $ 0.75\pm 0.11$ & $ 3.65\pm 0.59$ & $ 2.57\pm 1.05$ & $ 3.57\pm 0.15$ \\ 
   .042 &56288.89 & $ 0.31\pm 0.08$ & $ 1.03\pm 0.15$ & $ 0.85\pm 0.11$ & $ 4.74\pm 0.67$ & $ 2.26\pm 0.82$ & $ 3.70\pm 0.16$ \\ 
   .129 &56289.38 & $ 0.38\pm 0.13$ & $ 0.54\pm 0.10$ & $ 0.83\pm 0.11$ & $ 5.16\pm 0.96$ &   \nodata       & $ 3.67\pm 0.15$ \\

\enddata                                                                                                                          
\tablecomments{Null entries imply unresolved ratio and/or parameter ranges.}                                                      
\end{deluxetable}                                                                                                                 

%% file: TABLE_HFIR_O_6A.tex
                                                                                                                                  
\begin{deluxetable}{cccccccc}                                                                                                     
\tabletypesize{\footnotesize}                                                                                                     
                                                                                                                                  
\tablewidth{0pt}                                                                                                                  
\tablecolumns{8}                                                                                                                  
                                                                                                                                  
\tablecaption{Oxygen Line Ratios and Derived Parameters \label{tab:FIO}}                                                          
\tablehead{                                                                                                                       
\colhead{phase}          &                                                                                                        
\colhead{MJD}            &                                                                                                        
\colhead{$f/i$ ratio}    &                                                                                                        
\colhead{$G$ ratio}      &                                                                                                        
\colhead{$H/He$ ratio}   &                                                                                                        
\colhead{$R_{fir}/R{*}$} &                                                                                                        
\colhead{$T_{G}$  MK}    &                                                                                                        
\colhead{$T_{HHe}$  MK}}                                                                                                          
\startdata                                                                                                                        
                                                                                                                                  
\multicolumn{8}{c}{\textit{\underline{Time Ordered}}} \\ \\                                                                       
                                                                                                                                  
   .646 &56280.93 & $ 0.12\pm 0.08$ & $ 0.84\pm 0.20$ & $ 1.05\pm 0.19$ & $ 9.05\pm 3.40$ & $ 3.40\pm 1.90$ & $ 2.32\pm 0.13$ \\ 
   .734 &56281.38 & $ 0.06\pm 0.05$ & $ 2.50\pm 1.01$ & $ 2.55\pm 0.94$ & $ 6.29\pm 2.88$ & $ 0.00\pm 0.62$ & $ 2.98\pm 0.36$ \\ 
   .777 &56281.84 & $ < 0.08$       & $ 0.96\pm 0.27$ & $ 1.26\pm 0.31$ & $ < 6.9$        & $ 2.70\pm 1.63$ & $ 2.43\pm 0.18$ \\ 
   .082 &56283.49 & $ < 0.02$       & $ 1.19\pm 0.34$ & $ 1.73\pm 0.39$ & $ < 3.8$        & $ 1.59\pm 0.99$ & $ 2.68\pm 0.17$ \\ 
   .170 &56283.97 & $ < 0.02$       & $ 0.83\pm 0.25$ & $ 1.50\pm 0.32$ & $ < 4.0$        & $ 4.73\pm 3.32$ & $ 2.56\pm 0.16$ \\ 
   .257 &56284.44 & $ < 0.02$       & $ 0.90\pm 0.27$ & $ 1.43\pm 0.31$ & $ < 4.1$        & $ 3.31\pm 2.14$ & $ 2.52\pm 0.16$ \\ 
   .475 &56285.79 & $ < 0.02$       & $ 0.74\pm 0.23$ & $ 1.17\pm 0.25$ & $ < 3.9$        & $ > 1.8$        & $ 2.39\pm 0.15$ \\ 
   .562 &56286.28 & $ < 0.04$       & $ 0.68\pm 0.18$ & $ 1.37\pm 0.26$ & $ < 5.3$        & $ > 2.5$        & $ 2.50\pm 0.14$ \\ 
   .649 &56286.77 & $ 0.09\pm 0.08$ & $ 0.36\pm 0.11$ & $ 0.79\pm 0.14$ & $ 6.52\pm 3.89$ & $ > 77$         & $ 2.15\pm 0.09$ \\ 
   .955 &56288.40 & $ < 0.08$       & $ 1.11\pm 0.30$ & $ 1.66\pm 0.35$ & $ < 7.6$        & $ 1.78\pm 1.07$ & $ 2.64\pm 0.16$ \\ 
   .042 &56288.89 & $ 0.05\pm 0.05$ & $ 0.97\pm 0.45$ & $ 1.60\pm 0.53$ & $ 5.26\pm 3.31$ & $ > 0.35$       & $ 2.59\pm 0.26$ \\ 
   .129 &56289.38 & $ < 0.04$       & $ 0.83\pm 0.22$ & $ 1.38\pm 0.27$ & $ < 7.2$        & $ 3.70\pm 2.22$ & $ 2.50\pm 0.14$ \\

\enddata                                                                                                                          
\end{deluxetable}